\documentclass[10pt, aps, prd, superscriptaddress, nofootinbib, showpacs, twocolumn]{revtex4-2}
\usepackage[english]{babel}
\usepackage{amsmath, amssymb, amsfonts, amsthm, mathrsfs}
\usepackage[all]{xy}
\usepackage{bm}
\usepackage{stmaryrd}
\usepackage{graphicx}
\usepackage[colorlinks]{hyperref}
\usepackage{color}

\makeatletter

\renewcommand*{\p@subsection}{}

\renewcommand*{\p@subsubsection}{}
\makeatother
\numberwithin{equation}{section}

\DeclareRobustCommand{\vect}[1]{\bm{#1}}
\newcommand{\bbC}{\mathbb C}

\newcommand{\calA}{\mathcal A}
\newcommand{\calC}{\mathcal C}
\newcommand{\calD}{\mathcal D}
\newcommand{\calK}{\mathcal K}
\newcommand{\calF}{\mathcal F}
\newcommand{\calJ}{\mathcal J}
\newcommand{\calL}{\mathcal L}

\newcommand{\scrA}{\mathscr A}
\newcommand{\scrF}{\mathscr F}

\newcommand{\scrX}{\mathscr X}
\newcommand{\scrY}{\mathscr Y}

\newcommand{\frakA}{\mathfrak{A}}
\newcommand{\frakP}{\mathfrak{P}}
\newcommand{\frakR}{\mathfrak{R}}

\DeclareMathOperator{\tr}{tr}
\DeclareMathOperator{\Hrm}{\mathsf{Hrm}}
\DeclareMathOperator{\aHrm}{\mathsf{aHrm}}

\begin{document}
% ======================================

\title{``Metric-affine-like'' generalization of YM (mal-YM): detailed classical consideration}

\author{W\l adys\l aw Wachowski}
\email{vladvakh@gmail.com}
\affiliation{Theory Department, Lebedev Physics Institute, Leninsky Prospect 53, Moscow 119991, Russia}

\begin{abstract}
We consider the ``metric-affine-like'' generalization of the Yang-Mills theory (mal-YM) which we first proposed earlier. In this model, the connection is no longer assumed to be compatible with the Hermitian form in the fibers. As a consequence, along with the usual YM potential $\bm{A}_a$ and the field strength tensor $\bm{F}_{ab}$, it contains non-trivially interacting fields $\bm{B}_a$, $\bm{h}$, and $\bm{G}_{ab}$, $\bm{N}_a$, forming a non-Abelian generalization of St\"{u}ckelberg theory. Due to the spontaneous symmetry breaking $GL(n,\mathbb{C}) \to U(n)$, these new fields can be made massive and the limit $M\to\infty$ restores the standard YM theory. We perform a detailed analysis of this theory on the classical level. We discuss in detail geometric motivation for the model, field transformations, gauge symmetry and its spontaneous breaking, action, equations of motion, Noether identities, gauge fixing, and other issues.
\end{abstract}

\maketitle
\begingroup
  \hypersetup{linkcolor=blue}
  \tableofcontents
\endgroup

%%%%%%%%%%%%%%%%%%%%%%%%%%%%%%%%%%%%%
\section{Introduction}
%%%%%%%%%%%%%%%%%%%%%%%%%%%%%%%%%%%%%

This article is devoted to a detailed consideration (for now, exclusively at the classical level) of a novel generalization of Yang-Mills (YM) theory, which we first proposed in \cite{Wachowski24a}.

It would be appropriate to lay all our cards on the table from the outset and briefly formulate the main idea and features of this generalization. Our starting point was the metric-affine gravity (MAG) \cite{Hehl1976, Buchbinder1992, Hehl1995, PBO2017, Baldazzi2022}---a generalization of ordinary Einstein gravity (EG), in which, however, the metric $g_{ab}$ is no longer assumed to be covariantly constant $\nabla_a g_{bc} \ne 0$. As a result, the connection $\Gamma_{ab}{}^c$ is no longer the Levi--Civita connection uniquely determined by the metric, but along with the metric constitutes an independent dynamical field. Our key observation is that YM theory with a gauge group $G$ always implies some additional structure in fibers of the vector bundle, which is preserved by transformations from $G$. In particular, if we consider a theory with group $\mathrm{U}(n)$, as we will do throughout this paper for simplicity, then this structure is a Hermitian form $g_{\alpha\beta'}$ (the primed index notation will be explained in the main text of the paper). This structure in the fibers---in our case $g_{\alpha\beta'}$---is a certain analogue of the metric $g_{ab}$ in EG. In particular, in YM it is always assumed that $g_{\alpha\beta'}$ is such that the connection is compatible with it
\begin{equation} \label{CovConstYM}
\nabla_a g_{\alpha\beta'} = 0.
\end{equation}
A consequence of this condition is that the potential $\bm{A}_a$ and the strength tensor $\bm{F}_{ab}$ in YM take values in the Lie algebra of the corresponding group $G$, i.e., in our $\mathrm{U}(n)$ case they become anti-Hermitian $(n\times n)$-matrices.\footnote{Or Hermitian---it is just a matter of convention. It is known that in the complex case, Hermitian and anti-Hermitian matrices differ from one another by a factor of $i$.} Accordingly, we asked the following simple question: what happens if, by analogy with MAG, we abandon the condition \eqref{CovConstYM}, thus treating the connection and the structure in the fibers as two independent dynamical variables? Due to this analogy, we call this generalization of YM \emph{``metric-affine-like''}---mal-YM.

It turns out that mal-YM can not only be used as a toy model for better understanding the properties of MAG, but also has very interesting properties in its own right. We also believe that it has the potential to be significant as a real theory of fundamental interactions. Abandoning the condition~\eqref{CovConstYM} results in the total potential $\calA_a$ and the total curvature in the bundle $\calF_{ab}$ no longer being anti-Hermitian, but rather arbitrary complex matrices describing the general linear connection. Our theory (like any other with $n$-dimensional complex internal space) initially possesses the almost trivial $\mathrm{GL}(n, \bbC)$ gauge symmetry, and the Hermitian form $g_{\alpha\beta'}$ is a Higgs-like field that spontaneously breaks this symmetry to $\mathrm{U}(n)$. Using $g_{\alpha\beta'}$, we can split $\calA_a$ and $\calF_{ab}$ into Hermitian and anti-Hermitian parts---thus, along with the usual anti-Hermitian fields $\bm{A}_a$ and $\bm{F}_{ab}$, mal-YM acquires their Hermitian counterparts $\bm{B}_a$ and $\bm{G}_{ab}$, non-trivially interacting with them. And, due to spontaneous symmetry breaking, these fields can be assigned an arbitrary mass $M$.

A small perturbation of the Hermitian form $g_{\alpha\beta'}$---the Hermitian matrix $\bm{h}$---is a Goldstone boson, which together with the field $\bm{B}_a$ forms a non-Abelian generalization of the St\"{u}ckelberg theory \cite{Ruegg2004}. Using residual non-unitary transformations from $\mathrm{GL}(n, \bbC)$, the field $\bm{h}$ can be completely eliminated (this is an analogue of choosing the unitary gauge in the Higgs mechanism), in which case $\bm{B}_a$ becomes a massive Proca field whose propagator does not decrease in the UV region. However, in the Feynman-'t~Hooft gauge, this difficulty does not arise: the propagators of both fields $\bm{B}_a$ and $\bm{h}$ behave well, which gives hope for the renormalizability of mal-YM. The price for this is that $\bm{h}$ becomes a dynamical scalar field with mass $M$, interacting non-polynomially with the fields $\bm{A}_a$ and $\bm{B}_a$. Importantly, as the mass of the St\"{u}ckelberg sector fields tends to infinity $M\to\infty$, they are effectively frozen out, and the standard YM theory is restored.

All these statements will be explained and justified in detail in the main text of the paper. It is structured as follows: in the introductory Sec.~\ref{ConnectionSec}, we discuss in detail the general linear connection and introduce the total potential $\calA_a$ and the total curvature in the bundle $\calF_{ab}$. Since we will further use the background field formalism, we will do so not as usual, but in an algebraic way, in the spirit of~\cite{PenroseRindler1}; to emphasize the analogy with gravity, and since this requires no additional effort, we will simultaneously present the corresponding expressions for an affine connection. At the end of the section, we will also explain why every field theory with $n$-dimensional complex interior space has $\mathrm{GL}(n, \bbC)$ gauge symmetry. Next, in Sec.~\ref{HermitianFormSec}, we introduce the Hermitian form $g_{\alpha\beta'}$, define the operation of Hermitian conjugation on matrices, and decompose $\calA_a$ and $\calF_{ab}$ into its Hermitian and anti-Hermitian parts $\bm{A}_a$, $\bm{B}_a$, $\bm{F}_{ab}$, and $\bm{G}_{ab}$. We no longer assume that the condition \eqref{CovConstYM} is satisfied and introduce the YM-deviation vector $\bm{N}_a$ as a measure of its violation. Importantly, the covariant derivative now ceases to commute with Hermitian conjugation, and the Hermitian part of the total curvature $\bm{G}_{ab}$ is then uniquely expressed in terms of $\bm{N}_a$. In this section, we also discuss in detail the transformations of all quantities of our theory with respect to variation of the connection and the Hermitian form. Sec.~\ref{GaugeSec} is devoted to a discussion of gauge symmetry: we consider the spontaneous symmetry breaking $\mathrm{GL}(n, \bbC) \to \mathrm{U}(n)$, treat all emerging fields within the background field formalism, then introduce the sources of the fields and derive Noether identities for them.

Following all these preparations, in Sec.~\ref{ActionSec}, we introduce the mal-YM action. From it, we derive and analyze nonlinear equations of motion (EoMs) for background fields and linearized EoMs for small perturbations on a trivial background. At the end of this section, we perform the standard redefinition of fields and gauge couplings. Of conceptual importance is Sec.~\ref{GaugeFixingSec}, devoted to gauge fixing. In it we show that although the Goldstone boson $\bm{h}$ can be completely eliminated in the ``unitary'' gauge, this is not a good choice, since it leads to a non-decreasing propagator of the massive field $\bm{B}_a$. Therefore, it is more convenient to consider the Feynman-'t~Hooft gauge, which does not lead to problems with propagator behavior, but leads to nonpolynomial interactions with the field $\bm{h}$. Interactions in a trivial background are considered in Sec.~\ref{InteractionsSec}, where we first note that mal-YM has some additional discrete symmetry and an associated conserved parity-like quantity. Then we obtain vertices of pure mal-YM on a trivial background and, finally, consider the interaction with a gauge-charged scalar. In Conclusion~\ref{ConclusionSec}, we discuss the potential significance of this theory, primarily the open question of its renormalizability, as well as its possible applications in particle physics. Two appendices contain some auxiliary material: in Appendix~\ref{CurvApp} we discuss a general linear connection, and provide expressions for a nontrivial background in Appendix~\ref{VertApp}.
\newpage
%%%%%%%%%%%%%%%%%%%%%%%%%%%%%%%%%%%%%
\section{Connection $\nabla_a$} \label{ConnectionSec}
%%%%%%%%%%%%%%%%%%%%%%%%%%%%%%%%%%%%%

\subsection{Connection and potentials}

First of all, especially in light of the further use of the background field method, it will be convenient for us to follow Penrose and Rindler \cite{PenroseRindler1} and slightly change our view of the connection. In the traditional approach, the action of the covariant derivative $\nabla_a$ on tensors is defined via the vector potential $\bm{A}_a \cong \scrA_{a\beta}{}^\alpha$ and the Christoffel symbols $\Gamma_{ac}{}^b$ according to the relations
\begin{align}
\nabla_a \bm{\psi} = (\partial_a - ie\bm{A}_a) \bm{\psi}, \label{CovDerYM} \\
\nabla_a v^b = \partial_a v^b + \Gamma_{ac}{}^b v^c, \label{SpacetimeCovDer}
\end{align}
where $\bm{\psi} \cong \psi^\alpha$ is a gauge-charged scalar field.\footnote{Commonly used expressions like $\bm{\psi} = \psi^\alpha$ are an abuse of notation, since they immediately lead to absurd consequences, for example, $\psi^\alpha = \psi^\beta$, $\psi^\alpha\chi^\beta - \psi^\beta\chi^\alpha = 0$, etc. Therefore, we prefer to connect expressions obtained from each other by discarding or renaming indices, with the isomorphism sign $\cong$. Then $\bm{\psi} \cong \psi^\alpha \cong \psi^\beta \cong \ldots$, while $\psi^\alpha\chi^\beta - \psi^\beta\chi^\alpha \not\cong 0$. In our consideration, we do not use the choice of basis anywhere, so all indices should be understood as abstract. But, if it is more convenient for the reader, one can understand them in the usual sense as components, relative to the chosen basis.} (We denote spacetime indices everywhere with lowercase Latin letters, and internal color indices with lowercase Greek letters) and $e$ is the gauge coupling. It will be convenient for us to reverse this logic, considering the covariant derivative $\nabla_a$ (which we identify throughout with the connection) as the original fundamental object, defined not via \eqref{CovDerYM}-\eqref{SpacetimeCovDer}, but purely algebraically.

Namely, we define the covariant derivative $\nabla_a$ as an operation on tensors $U^\scrX$ (where $\scrX$ is a symbolic notation for an arbitrary set of spacetime and internal color indices) that adds one spacetime subscript $a$, coincides on scalars with the canonically defined differential $df \cong \nabla_a f$, acts additively
\begin{equation}
\nabla_a (U^\scrX + V^\scrX) = \nabla_a U^\scrX + \nabla_a V^\scrX,
\end{equation}
and acts on tensor products according to the Leibniz rule
\begin{equation} \label{LeibnizRule}
\nabla_a (U^\scrX V^\scrY) = V^\scrY \nabla_a U^\scrX + U^\scrX \nabla_a V^\scrY.
\end{equation}
The listed properties fully characterize the connection $\nabla_a$ and do not yet contain any reference to the potential $\bm{A}_a$ or the Christoffel symbols $\Gamma_{ac}{}^b$.

Now consider the following question: if we are given two different connections $\nabla_a$ and $\tilde\nabla_a$, how are they related to each other? How can we describe their difference? Since they coincide on scalars $(\tilde\nabla_a - \nabla_a) f = 0$, their difference acts linearly on tensors
\begin{multline}
(\tilde\nabla_a - \nabla_a) (fU^\scrX + gV^\scrX) \\
= f(\tilde\nabla_a - \nabla_a)U^\scrX + g(\tilde\nabla_a - \nabla_a)V^\scrX,
\end{multline}
and therefore must itself be described by some tensors. In fact, the Leibniz rule implies that only one tensor is needed for each type of indices: if we define
\begin{align}
(\tilde\nabla_a - \nabla_a)V^b &= \scrA_{ac}{}^b V^c, \label{GaugePotencialDef2} \\
(\tilde\nabla_a - \nabla_a)\varphi^\alpha &= \scrA_{a\beta}{}^\alpha \varphi^\beta, \label{GaugePotencialDef}
\end{align}
then the action of the difference $\tilde\nabla_a - \nabla_a$ on an arbitrary tensor $H^{b\ldots\beta\ldots}_{c\ldots\alpha\ldots}$ is given by
\begin{align} \label{RaznDiff}
(\tilde\nabla_a &- \nabla_a) H^{b\ldots\beta\ldots}_{c\ldots\alpha\ldots} \\
&= \scrA_{ab_0}{}^b H^{b_0\ldots\beta\ldots}_{c\ldots\alpha\ldots} + \ldots - \scrA_{ac}{}^{c_0} H^{b\ldots\beta\ldots}_{c_0\ldots\alpha\ldots} - \ldots \nonumber \\
&+ \scrA_{a\beta_0}{}^\beta H^{b\ldots\beta_0\ldots}_{c\ldots\alpha\ldots} + \ldots - \scrA_{a\alpha}{}^{\alpha_0} H^{b\ldots\beta\ldots}_{c\ldots\alpha_0\ldots} - \ldots. \nonumber
\end{align}
It is fundamentally important that \emph{connection potentials} $\frakA_a[\tilde\nabla-\nabla] \cong \scrA_{ac}{}^b$ and $\calA_a[\tilde\nabla-\nabla] \cong \scrA_{a\beta}{}^\alpha$\footnote{To avoid confusion, we will write internal color space matrices in bold calligraphic font, and tangent space matrices in Gothic font.} depend \emph{not on one, but on two connections}, both on $\tilde\nabla_a$ and on $\nabla_a$. It is easy to see that the converse is also true: if $\nabla_a$ is some covariant derivative, then the operation $\tilde\nabla_a$ defined according to \eqref{RaznDiff}, where $\frakA_a$ and $\calA_a$ are arbitrary tensors, will also be a covariant derivative.

Thus, \emph{the space of all connections has the structure of an affine space}. Recall that an affine space differs from a vector space in that it does not have a distinguished origin. Therefore in an affine space we can no longer add its points or multiply them by a scalar, as we can do with vectors in a vector space. However, for each two points $x$ and $y$ of an affine space, the vector of their difference $\bm{v} = y-x$ is defined, starting at the point $x$ and ending at $y$. And vice versa, for each point $x$ and each vector $\bm{v}$, a new point $y = x + \bm{v}$ is defined, obtained by shifting from $x$ by $\bm{v}$. The situation with connections is absolutely analogous: in the space of all connections there is no distinguished canonical zero element, we cannot add connections or multiply them by a scalar, but the difference of two connections $\tilde\nabla_a - \nabla_a$ is determined by two tensors $\frakA_a[\tilde\nabla-\nabla]$ and $\calA_a[\tilde\nabla-\nabla]$.

How do the potentials $\frakA_a[\tilde\nabla-\nabla]$ and $\calA_a[\tilde\nabla-\nabla]$ that we have introduced relate to the traditional $\bm{A}_a$ and $\Gamma_{ac}{}^b$ in the relations \eqref{CovDerYM}-\eqref{SpacetimeCovDer}? The answer is that these relations implicitly assume that not only the connection $\nabla_a$ under consideration is given, but also some second auxiliary connection $\partial_a$, which can be called ``zero'' or ``basic'' and which is determined by differentiations with respect to a given coordinate system. Comparing the expressions \eqref{CovDerYM}-\eqref{SpacetimeCovDer} with the definitions \eqref{GaugePotencialDef2}-\eqref{GaugePotencialDef}, we obtain
\begin{equation}
- ie\bm{A}_a[\nabla] = \calA_a[\nabla-\partial], \quad \Gamma_{ac}{}^b[\nabla] = \scrA_{ac}{}^b[\nabla-\partial].
\end{equation}
The inverse relations are obviously of the form
\begin{align}
\calA_a[\tilde\nabla-\nabla] &= -ie(\bm{A}_a[\tilde\nabla] - \bm{A}_a[\nabla]), \\
\scrA_{ac}{}^b[\tilde\nabla-\nabla] &= \Gamma_{ac}{}^b[\tilde\nabla] - \Gamma_{ac}{}^b[\nabla].
\end{align}
In other words, what we call ``potentials'' for brevity in this paper are actually variations of the commonly used vector potentials and Christoffel symbols.

It is important to note that as long as we do not impose any additional conditions on the connection related to the presence of additional structures in the tangent or internal color space, the potentials $\frakA_a[\tilde\nabla-\nabla]$ and $\calA_a[\tilde\nabla-\nabla]$ are arbitrary tensors without any additional symmetries. In particular, $\calA_a[\tilde\nabla-\nabla]$ is not assumed to be an anti-Hermitian matrix here.

It is also worth noting that the well-known assertion that ``Christoffel symbols are not tensors'' is simply due to the fact that $\Gamma_{ac}{}^b$ implicitly depends on the ``basic'' connection $\partial_a$. Therefore, a change of coordinates leads to its change and the components of $\Gamma_{ac}{}^b$ are transformed differently than the components of an ordinary tensor. However, the potentials $\frakA_a[\tilde\nabla-\nabla]$ and $\calA_a[\tilde\nabla-\nabla]$ do not depend on the ``basic'' connection $\partial_a$ at all. Therefore, when changing coordinates with fixed connections $\tilde\nabla_a$ and $\nabla_a$, their components are transformed in the usual way. So potentials $\frakA_a[\tilde\nabla-\nabla]$ and $\calA_a[\tilde\nabla-\nabla]$ are genuine tensors.

The use of potentials $\frakA_a[\tilde\nabla-\nabla]$ and $\calA_a[\tilde\nabla-\nabla]$ is convenient in the sense that it does not imply the choice of any ``basic'' connection $\partial_a$, coordinate system, etc. and is therefore free from the arbitrariness associated with this. So it corresponds much better to the intrinsic structure of connection. It seems especially natural in the context of the background field method (see Subsection \ref{BackgroundSubsec}). In this case, we assume that among all possible connections there is some distinguished physical connection $\nabla_a$ that carries degrees of freedom. We regard this physical connection as a fixed \emph{background field}. When we need to vary with respect to connection, we pass from $\nabla_a$ to some new connection $\tilde\nabla_a$, which differs from the physical one by the potentials $\frakA_a[\tilde\nabla-\nabla]$ and $\calA_a[\tilde\nabla-\nabla]$, and we will consider the change of all connection-dependent quantities caused by this transformation. Thus, the potentials $\frakA_a[\tilde\nabla-\nabla]$ and $\calA_a[\tilde\nabla-\nabla]$ will be just \emph{perturbations} of the connection components. This partition into background fields and perturbations is precisely the essence of the background field method, while the ``basic'' connection $\partial_a$ becomes superfluous and does not participate in any way in the consideration.

%%%%%%%%%%%%%%%%%%%%%%%%%%%%%%%%%%%%%%%%%%%
\subsection{Curvatures}

The next task is to learn how to commute covariant derivatives, which is done by introducing torsion and curvatures.

First of all, we note that due to the cancellation of the cross terms, the commutator of covariant derivatives $[\nabla_a, \nabla_b]$ acts on the product of scalars according to the Leibniz rule
\begin{equation}
[\nabla_a, \nabla_b](fg) = g[\nabla_a, \nabla_b]f + f [\nabla_a, \nabla_b]g,
\end{equation}
from which it follows that it must be linearly related to their differentiation
\begin{equation} \label{DefT}
[\nabla_a, \nabla_b] f = T_{ab}{}^c[\nabla] \nabla_c f,
\end{equation}
where the antisymmetric tensor $T_{ab}{}^c[\nabla] = T_{[ab]}{}^c$ is called \emph{torsion}.

Let us repeat this reasoning once more, now not for scalars, but for arbitrary tensors. To do this, consider the operator
\begin{equation} \label{DeltaCommOperator}
\Delta_{ab} = [\nabla_a, \nabla_b] - T_{ab}{}^c \nabla_c.
\end{equation}
It is easy to check that it also acts on tensor products according to the Leibniz rule
\begin{equation} \label{DeltaLeibnizRule}
\Delta_{ab}(U^\scrX V^\scrY) = V^\scrY\Delta_{ab} U^\scrX + U^\scrX\Delta_{ab} V^\scrY.
\end{equation}
Moreover, by definition \eqref{DefT} $\Delta_{ab}$ vanishes on scalars and, therefore, acts linearly on tensors and is determined itself by some tensors. Then again, by virtue of the Leibniz rule \eqref{DeltaLeibnizRule}, we can conclude that its action on arbitrary tensors must be determined by one tensor for each type of indices. Namely, if we define
\begin{gather}
\Delta_{ab} V^d = R_{abc}{}^d[\nabla] V^c, \label{DefR} \\
\Delta_{ab} \psi^\alpha = \scrF_{ab\beta}{}^\alpha[\nabla] \psi^\beta, \label{DefF}
\end{gather}
then the action of the operator $\Delta_{ab}$ on an arbitrary tensor $H^{c\ldots\beta\ldots}_{d\ldots\alpha\ldots}$ will be given by a relation completely analogous to \eqref{RaznDiff}---the so-called generalized Ricci identity:
\begin{align}
&\Delta_{ab} H^{c\ldots\beta\ldots}_{d\ldots\alpha\ldots} \label{GenRicci} \\
&= R_{abc_0}{}^c H^{c_0\ldots\beta\ldots}_{d\ldots\alpha\ldots} + \ldots - R_{abd}{}^{d_0} H^{c\ldots\beta\ldots}_{d_0\ldots\alpha\ldots} - \ldots \nonumber \\
&+ \scrF_{ab\beta_0}{}^\beta H^{c\ldots\beta_0\ldots}_{d\ldots\alpha\ldots} + \ldots - \scrF_{ab\alpha}{}^{\alpha_0} H^{c\ldots\beta\ldots}_{d\ldots\alpha_0\ldots} - \ldots. \nonumber
\end{align}
The tensors $\frakR_{ab}[\nabla] = \frakR_{[ab]} \cong R_{abc}{}^d$ and $\calF_{ab}[\nabla] = \calF_{[ab]} \cong \scrF_{ab\beta}{}^\alpha$, which are antisymmetric with respect to the first pair of indices, are called, respectively, the \emph{spacetime curvature} and the \emph{curvature in the bundle}.

An important difference between the torsion $T_{ab}{}^c[\nabla]$ and the curvatures $\frakR_{ab}[\nabla]$, $\calF_{ab}[\nabla]$ on the one hand and the potentials $\frakA_a[\tilde\nabla-\nabla]$, $\calA_a[\tilde\nabla-\nabla]$ on the other is as follows: potentials always depend on \emph{two} connections $\tilde\nabla_a$ and $\nabla_a$, but the torsion and curvatures depend on a \emph{single} connection $\nabla_a$. These are fundamental geometric characteristics from which, along with the covariant derivative $\nabla_a$ itself, any geometrically defined action must be constructed. Therefore, it is necessary to find how they are transformed when we pass from the connection $\nabla_a$ to some other connection $\tilde\nabla_a$.

First of all, it is easy to show (see Appendix \ref{CurvApp}) that the torsion transforms as follows:
\begin{equation} \label{TorsionPreobr}
T_{ab}{}^c[\tilde\nabla] - T_{ab}{}^c[\nabla] = - 2\scrA_{[ab]}{}^c.
\end{equation}
Therefore, if we are given a connection $\nabla_a$ with nonzero torsion $T_{ab}{}^c[\nabla] \ne 0$, we can always pass to a new connection $\tilde\nabla_a$, for which there is no torsion $T_{ab}{}^c[\tilde\nabla] = 0$, using the transformation
\begin{equation} \label{ToSymmetric}
\scrA_{ab}{}^c[\tilde\nabla-\nabla] = \frac{1}{2} T_{ab}{}^c[\nabla].
\end{equation}
Such a connection is called symmetric or torsionless. Therefore, everywhere below, as is usually done, we will assume this transition has been performed and consider only torsionless connections.

Further, if there is no torsion, the curvature transformation laws have the form
\begin{align}
&\calF_{ab}[\tilde\nabla] - \calF_{ab}[\nabla] = \nabla_a\calA_b - \nabla_b\calA_a + [\calA_a, \calA_b], \label{FPreobr} \\
&\frakR_{ab}[\tilde\nabla] - \frakR_{ab}[\nabla] = \nabla_a\frakA_b - \nabla_b\frakA_a + [\frakA_a, \frakA_b], \label{tildeFPreobr}
\end{align}
where the right-hand side contains the potentials $\frakA_a[\tilde\nabla-\nabla]$ and $\calA_a[\tilde\nabla-\nabla]$.

Also it can be shown that the curvature tensors satisfy the algebraic Bianchi identity
\begin{equation} \label{AlgBianSym}
R_{[abc]}{}^d = 0,
\end{equation}
and differential Bianchi identities
\begin{equation} \label{DifBian}
\nabla_{[a} \calF_{bc]} = 0,  \qquad \nabla_{[a} \frakR_{bc]} = 0.
\end{equation}
The derivation of the relations \eqref{FPreobr}-\eqref{DifBian} together with their generalizations for the case of nonzero torsion $T_{ab}{}^c[\nabla] \ne 0$ is given in Appendix \ref{CurvApp} for reference.

Note that the compact and elegant formalism of matrix-valued differential forms is extremely convenient for working with connection, potentials, and curvatures. We will not introduce the corresponding notations in the main text of this paper to make it more understandable, but for reference we will give them in Subsection~\ref{FormApp} of Appendix~\ref{CurvApp}.

We emphasize again that without additional structures in the fibers and the additional conditions on the connection associated with them, the Bianchi identities and antisymmetry with respect to the first two indices are the only restrictions on the form of the curvature tensors $\calF_{ab}$ and $\frakR_{ab}$. In particular, the tensor $\frakR_{ab}$ is not antisymmetric with respect to the matrix indices, and the tensor $\calF_{ab}$ is not anti-Hermitian, but an arbitrary complex matrix (without a metric $g_{ab}$ and a Hermitian form $g_{\alpha\beta'}$, it is impossible even to introduce the corresponding concepts of antisymmetry and anti-hermiticity of an object with upper and lower indices).

In order to emphasize the deep kinship between YM and the theory of gravity and the common geometric structures underlying them, so far we have conducted our presentation in parallel for the spacetime ($\frakA_a$, $\frakR_{ab}$) and internal color ($\calA_{a}$, $\calF_{ab}$) components of the connection. It is worth noting that the approach considered here allows in principle to construct YM on the most general curved background (including an metric-affine one). However, at the present moment our task is to construct mal-YM and to clarify its key features. In order to do this in the simplest and most clear way, it is convenient to get rid of all cumbersome and obscure complications. Therefore, since the curvature transformations \eqref{FPreobr} and \eqref{tildeFPreobr} are independent of each other, from now until the end of the paper we will consider the case when the spacetime is flat, i.e. we consider only such physical connections $\nabla_a$ for which the Riemann tensor vanishes $R_{abc}{}^d[\nabla]=0$ (and the metric is Euclidean $g_{ab} = \delta_{ab}$). The generalization to the general curved case is straightforward, and we will not touch upon these issues here.

%%%%%%%%%%%%%%%%%%%%%%%%%
\subsection{General $\mathrm{GL}(n,\bbC)$ gauge symmetry} \label{InvTransformSect}

Let us agree to call linear maps of the internal color space $V\to V$ ``matrices'', i.e. ``matrices'' are tensors with two color indices---one upper and one lower $\bm{M} \cong M_\alpha^\beta$, so that the product of matrices is simply a contraction with respect to one of the indices $\bm{M}\bm{N} \cong M^\alpha_\gamma N^\gamma_\beta$.

Let $\bm{u} \cong u_\alpha^\beta$ and $\bm{U} \cong U_\alpha^\beta$ be two arbitrary mutually inverse matrices:
\begin{equation}
\bm{u}\bm{U} = \bm{U}\bm{u} = \mathbf{1}.
\end{equation}
Then differentiating the invertibility condition yields:
\begin{equation} \label{InvMatrDif}
\nabla_a\bm{U} = -\bm{U} (\nabla_a\bm{u}) \bm{U}.
\end{equation}

Consider invertible linear transformations of the internal color space. In this transformation, all tensors with internal color indices are transformed into some new ones, where each upper color index is contracted with the matrix $\bm{U}$, and each lower one is contracted with the inverse matrix $\bm{u}$:
\begin{align} \label{InvColorTransform}
H_{\alpha_1\ldots \alpha_p}^{\beta_1\ldots\beta_q}{}^{\scrX} \mapsto& \tilde H_{\alpha_1\ldots \alpha_p}^{\beta_1\ldots\beta_q}{}^{\scrX} \nonumber
\\=U_{\gamma_1}^{\beta_1}\cdots U_{\gamma_q}^{\beta_q}\; &H_{\delta_1\ldots \delta_p}^{\gamma_1\ldots\gamma_q}{}^{\scrX}\; u_{\alpha_1}^{\delta_1} \cdots u_{\alpha_p}^{\delta_p}.
\end{align}
In particular, for a matrix $\bm{M}$ we will have
\begin{equation} \label{MatrixTrans}
\tilde{\bm M} = \bm{UMu}.
\end{equation}

It is clear that such a transformation will preserve all contractions, for example:
\begin{equation}
\tilde\chi_\gamma \tilde\psi^\gamma = \chi_\alpha u_\gamma^\alpha U_\beta^\gamma \psi^\beta = \chi_\gamma \psi^\gamma.
\end{equation}
But then, under the transformations \eqref{InvColorTransform}, any covariant scalar constructed from contractions of gauge-charged fields (i.e. fields carrying internal color indices) will be preserved, including, of course, the theory Lagrangian. In other words, the transformations under which all fields of the theory without exception are transformed according to \eqref{InvColorTransform} \emph{will always be gauge symmetry of this theory}\footnote{Note that the theory may also have other symmetries not described by these transformations.}.

However, in order for the transformation law~\eqref{InvColorTransform} to hold for all tensors without exception, it is necessary that, for example, the tensor $\nabla_a \bm{\psi}$ be transformed in exactly the same way as the tensor $\bm{\psi} \cong \psi^\alpha$. This can only happen if the covariant derivative is transformed simultaneously $\nabla_a \mapsto \tilde\nabla_a$. We have
\begin{align}
\nabla_a \bm{\psi} \mapsto \tilde\nabla_a \tilde{\bm\psi} &= \bm{U} \nabla_a{\bm\psi} = \nabla_a \tilde{\bm\psi} + \calA_a \tilde{\bm\psi} \nonumber \\
&= \bm{U} \nabla_a{\bm\psi} + (\nabla_a\bm{U}){\bm\psi} + \calA_a \bm{U}{\bm\psi},
\end{align}
from which we directly obtain an expression for the connection potential
\begin{equation} \label{PureGaugePotencalEq}
\calA_a[\tilde\nabla-\nabla] = -(\nabla_a\bm{U})\bm{u} = \bm{U} \nabla_a\bm{u}.
\end{equation}

It is easy to check that this transformation is consistent, i.e. that covariant derivatives of tensors of other types, such as $\nabla_a\chi_\alpha$, will also transform correctly. Moreover, the curvature in the bundle is transformed in exactly the same way:
\begin{align}
\calF_{ab}[\tilde\nabla] &= \calF_{ab}[\nabla] + 2\nabla_{[a}\calA_{b]} + 2\calA_{[a} \calA_{b]} \nonumber \\
&= \calF_{ab}[\nabla] + 2\nabla_{[a}(\bm{U} \nabla_{b]}\bm{u}) + 2(\bm{U} \nabla_{[a}\bm{u}) (\bm{U} \nabla_{b]}\bm{u}) \nonumber \\
&= \calF_{ab}[\nabla] + 2\bm{U} \nabla_{[a}\nabla_{b]}\bm{u} \nonumber \\
&+ 2(\nabla_{[a}\bm{U}) \nabla_{b]}\bm{u} + 2\bm{U} (\nabla_{[a}\bm{u})\bm{U} \nabla_{b]}\bm{u} \nonumber \\
&= \calF_{ab}[\nabla] + \bm{U} [\calF_{ab}[\nabla], \bm{u}] = \bm{U} \calF_{ab}[\nabla] \bm{u}.
\end{align}
Here we first use the formula for the transformation of the curvature \eqref{FPreobr}, then substitute into it the expression for the potential \eqref{PureGaugePotencalEq}. Using the relation \eqref{InvMatrDif} the two terms in the forth row cancel each other out, and we write out the second term in the third row using the Ricci identity \eqref{GenRicci}.

Let us represent the transformation matrices in the form
\begin{equation} \label{InfinitesimalTransforms}
\bm{u} = \exp(\bm{\epsilon}), \qquad \bm{U} = \exp(-\bm{\epsilon}),
\end{equation}
where $\bm{\epsilon} \cong \epsilon_\alpha^\beta$ is a completely arbitrary matrix, and the matrix exponential is understood in the sense of its (everywhere convergent) Taylor series:
\begin{equation} \label{TaylorSeries}
\exp(\bm{\epsilon}) = \sum\limits_{k=0}^\infty \frac{\bm{\epsilon}^k}{k!}.
\end{equation}
Then, discarding higher-order terms in $\bm{\epsilon}$, we can write the gauge transformations \eqref{MatrixTrans} and \eqref{PureGaugePotencalEq} in infinitesimal form:
\begin{equation} \label{InfTensTrans}
\calA_{a}[\tilde\nabla-\nabla] = \nabla_a\bm{\epsilon}, \qquad \delta\bm{M} = \tilde{\bm M} - \bm{M} = [\bm{M}, \bm{\epsilon}].
\end{equation}

%%%%%%%%%%%%%%%%%%%%%%%%%%%%%%%%%%%%%
\section{+ Hermitian form $g_{\alpha\beta'}$} \label{HermitianFormSec}
%%%%%%%%%%%%%%%%%%%%%%%%%%%%%%%%%%%%%

In the previous section we considered exclusively the general linear connection $\nabla_a$. In this section we investigate what happens when we introduce an additional structure in the fibers of the bundle. It seems most natural to begin with the unitary case---perhaps it is also distinguished from the point of view of quantum theory. In any case, for the sake of simplicity of presentation and to better emphasize our main idea, in this article we restrict ourselves to considering only a theory with $\mathrm{U}(n)$ symmetry.

In this case, the $n$-dimensional internal color space $V$ (a typical fiber of the bundle) will be complex. Complex conjugation will antilinearly map $V$ into \emph{another space} $\bar V$, complex conjugate to it. We will denote the indices associated with them by primes. Unprimed and primed indices are analogous to the undotted and dotted indices standardly used for Weyl spinors. Since they belong to different spaces $V$ and $\bar V$, they cannot be contracted with each other. However, complex conjugation turns primed indices into unprimed ones and vice versa:
\begin{equation}
\overline{H_{\alpha\ldots\beta'\ldots}^{\gamma\ldots\delta'\ldots}} = \bar H_{\alpha'\ldots\beta\ldots}^{\gamma'\ldots\delta\ldots}
\end{equation}

We assume everywhere that the covariant derivative is real in the sense that
\begin{equation}
\overline{\nabla_a \varphi^\alpha} = \nabla_a \bar\varphi^{\alpha'}.
\end{equation}
Then the complex conjugation of the relation \eqref{GaugePotencialDef} gives
\begin{equation}
(\tilde\nabla_a - \nabla_a) \bar\varphi^{\alpha'} = \bar\scrA_{a\beta'}{}^{\alpha'} \bar\varphi^{\beta'}.
\end{equation}
Thus, if under a connection change the unprimed index is contracted with the potential $\scrA_{a\beta}{}^\alpha$, then the primed index is contracted with the conjugate potential $\bar\scrA_{a\beta'}{}^{\alpha'}$.

Let a Hermitian form $g_{\alpha\beta'}$ be given in fibers, i.e. there is a tensor with two lower indices---primed and unprimed---that is real
\begin{equation} \label{gDef1}
\bar g_{\alpha\beta'} = g_{\alpha\beta'},
\end{equation}
and non-degenerate, i.e. there is an inverse tensor $g^{\alpha\beta'}$ such that
\begin{equation} \label{gDef2}
g_{\alpha\gamma'}g^{\beta\gamma'} = \delta_\alpha^\beta
\end{equation}
(the complex conjugation of the last relation gives $g_{\gamma\alpha'}g^{\gamma\beta'} = \delta_{\alpha'}^{\beta'}$). Due to the last property, we can lower and raise indices in the usual way. However, when changing the position, the unprimed index becomes primed and vice versa, for example, $\bar\varphi_\alpha = g_{\alpha\alpha'}\bar\varphi^{\alpha'}$. (Note that since primed and unprimed indices are indices of different types, they can be freely rearranged. But then for matrices in the unitary case it is only important which of the two indices is upper and which is lower, and the order in which they appear is unimportant. Therefore, we will always write matrix indices one above the other.)

The ``metric-affine-like'' generalization we have constructed is based on abandoning the covariant constancy condition \eqref{CovConstYM}. Then the points in the space of all possible classical field configurations will have the form $(\nabla_a, g_{\alpha\beta'})$, where the connection $\nabla_a$ and the Hermitian form $g_{\alpha\beta'}$ will be completely independent variables.

%%%%%%%%%%%%%%%%%%%%%%%%%%%%%%%%%%%%%%%%%%%%
\subsection{Covariant derivatives and hermiticity} \label{DerHermSec}

Let us note the following: if we have a matrix $\bm{M} \cong M_\alpha^\beta$, then the complex conjugate object $\bar M_{\alpha'}^{\beta'}$ is no longer a ``matrix'': it is a map of the complex conjugate space $\bar V\to \bar V$, it has primed indices, and therefore cannot be multiplied by ``matrices''. However, we can define the operation of \emph{Hermitian conjugation}, which maps matrices to matrices, but contains not only complex conjugation, but also ``transposition,'' i.e. contractions with $g_{\alpha\alpha'}$ and $g^{\beta\beta'}$:
\begin{equation} \label{HermConjDef}
\bm{M}^\dag \cong \bar M_\alpha^\beta = g_{\alpha\alpha'} \bar M_{\beta'}^{\alpha'} g^{\beta\beta'}.
\end{equation}
We emphasize that, although it is not explicitly stated in standard algebra courses, the common Hermitian conjugation always requires the specification of some Hermitian form $g_{\alpha\beta'}$.

This operation allows us to split any complex matrix into a Hermitian and an anti-Hermitian parts, which we will denote by the symbols $\Hrm$ and $\aHrm$:
\begin{align}
&\bm{M} = \bm{b} - i\bm{a}, \quad\text{where} \\
&\bm{b} = \Hrm\bm{M} = \frac{1}{2}(\bm{M} + \bm{M}^\dag), \\
&\bm{a} = \aHrm\bm{M} = \frac{i}{2}(\bm{M} - \bm{M}^\dag).
\end{align}
(Of course, both parts defined in this way will actually be Hermitian matrices: $\bm{b}^\dag = \bm{b}$ and $\bm{a}^\dag = \bm{a}$. This is the same as imaginary part of a complex number being a real number.)

However, the relaxation of the covariant constancy condition~\eqref{CovConstYM} leads to the appearance of a non-zero vector
\begin{equation} \label{YMdeviationDef}
\bm{N}_a \cong N_a{}_\alpha^\beta = -\frac{1}{2} g^{\beta\beta'} \nabla_a g_{\alpha\beta'},
\end{equation}
which, by definition, is Hermitian $\bm{N}_a^\dag = \bm{N}_a$. It is analogous to nonmetricity tensor in MAG, and we will call it the \emph{YM-deviation vector}.

This, in turn, leads to some rather unusual consequences. Obviously now the operations of covariant differentiation and raising/lowering indices cease to commute with each other. So we now have that, for example,
\begin{equation}
\nabla_a \psi_{\alpha'} \ne g_{\alpha\alpha'} \nabla_a \psi^\alpha.
\end{equation}
The unfamiliarity of this circumstance can easily lead to carelessness during differentiation and, accordingly, to errors. To avoid them, it is convenient to act as follows: for each tensor, from all possible arrangements of its indices, one of them should be considered canonical one (for example, for the total curvature, it can be the matrix arrangement $\calF_{ab} \cong \scrF_{ab}{}_\alpha^\beta$). Then before differentiation we should represent the differentiable expression as products of tensors in canonical form, differentiate it according to the Leibniz rule, and only then remove the extra factors $g_{\alpha\alpha'}$ and $g^{\beta\beta'}$ by raising or lowering corresponding indices.

A special case of this will be non-commutativity of covariant differentiation with the operation of Hermitian conjugation, which is expressed in the appearance of additional commutators with the YM-deviation vector $\bm{N}_a$. Indeed, for an arbitrary matrix $\bm{M}$ we have:
\begin{align}
\nabla_a\left(\bm{M}^\dag\right) &\cong \nabla_a \left(g_{\alpha\alpha'} \bar M_{\beta'}^{\alpha'} g^{\beta\beta'}\right) = g_{\alpha\alpha'}g^{\beta\beta'} \nabla_a \bar M_{\beta'}^{\alpha'} \nonumber \\
&+ \bar M^{\beta\alpha'}\nabla_a g_{\alpha\alpha'} + \bar M_{\alpha\beta'} \nabla_a g^{\beta\beta'} \nonumber \\
&\cong \left(\nabla\bm{M}\right)^\dag +2\left[\bm{N}_a, \bm{M}^\dag \right]. \label{DerHermEq}
\end{align}
Accordingly, now the (anti-)hermiticity of matrices is not preserved under differentiation---the covariant derivative of a Hermitian matrix will contain an anti-Hermitian part proportional to the commutator with $\bm{N}_a$.

Now, taking into account the remarks above, we split the total curvature in the bundle $\calF_{ab}$ \eqref{DefF} into Hermitian and anti-Hermitian parts:
\begin{align}
&\calF_{ab} = \bm{G}_{ab} - i\bm{F}_{ab}, \label{TotalCurvature} \\
&\bm{G}_{ab} = \Hrm\calF_{ab}, \qquad \bm{F}_{ab} = \aHrm\calF_{ab}. \label{FGdef}
\end{align}
Note that while the total curvature $\calF_{ab}[\nabla]$ depends only on the connection $\nabla_a$ and not on the Hermitian form $g_{\alpha\beta'}$, while its Hermitian $\bm{G}_{ab}[\nabla, g]$ and anti-Hermitian $\bm{F}_{ab}[\nabla, g]$ parts, as well as the YM-deviation vector $\bm{N}_a[\nabla, g]$, depend on both $\nabla_a$ and $g_{\alpha\beta'}$.

It turns out that the Hermitian component of curvature in the bundle $\bm{G}_{ab}$ is fully expressed through the YM-deviation vector $\bm{N}_a$. To obtain this fundamental relation, let us act on the Hermitian form $g_{\alpha\beta'}$ by the commutator of covariant derivatives and rewrite this expression, on the one hand, using the Ricci identity \eqref{GenRicci} and, on the other hand, by the definition \eqref{YMdeviationDef}
\begin{align}
[\nabla_a, \nabla_b] g_{\alpha\beta'} &= -\scrF_{ab}{}_\alpha^\gamma g_{\gamma\beta'} - \bar\scrF_{ab}{}_{\beta'}^{\gamma'} g_{\alpha\gamma'} = -2G_{ab\alpha\beta'} \nonumber \\
&= -2(\nabla_a N_{b\alpha\beta'} - \nabla_b N_{a\alpha\beta'}). \label{NtoG1}
\end{align}
Now it only remains, by raising the primed index, to bring this relation to matrix form:
\begin{align}
\bm{G}_{ab} &\cong G_{ab}{}_\alpha^\beta = \nabla_aN_b{}_\alpha^\beta - \nabla_bN_a{}_\alpha^\beta \nonumber \\
&+ g^{\beta\beta'}(N_b{}_\alpha^\gamma \nabla_a - N_a{}_\alpha^\gamma \nabla_b) g_{\gamma\beta'} \nonumber \\
&\cong \nabla_a\bm{N}_b - \nabla_b\bm{N}_a - 2\left[\bm{N}_a, \bm{N}_b\right]. \label{NtoG}
\end{align}
This relation means that in the standard YM theory, where $\bm{N}_a = 0$, the curvature in the bundle completely reduces to the usual YM field strength tensor $\calF_{ab} = -i\bm{F}_{ab}$.

%%%%%%%%%%%%%%%%%%%%%%%%%%%%%%%%%%%%%%%%%%%
\subsection{Hermitian form transformations}

A general transformation of the connection $\nabla_a \mapsto \tilde\nabla_a$ is described by the total potential $\calA_a[\tilde\nabla-\nabla]$ according to \eqref{RaznDiff}. At the same time we can perform a transformation of the Hermitian form $g_{\alpha\beta'} \mapsto \tilde g_{\alpha\beta'}$ (both should satisfy conditions \eqref{gDef1}-\eqref{gDef2}). What quantity describes this transformation?

From Hermitian forms $g_{\alpha\beta'}$ and $\tilde g_{\alpha\beta'}$, we construct the following matrices:
\begin{equation} \label{OmegaMatrices}
\bm{\omega} \cong \omega_\alpha^\beta = \tilde g_{\alpha\beta'} g^{\beta\beta'}, \qquad \bm{\Omega} \cong \Omega_\alpha^\beta = g_{\alpha\beta'} \tilde g^{\beta\beta'}.
\end{equation}
It is easy to verify that they are Hermitian $\bm{\omega}^\dag = \bm{\omega}$, $\bm{\Omega}^\dag = \bm{\Omega}$ and mutually inverse $\bm{\omega\Omega} = \bm{\Omega\omega} = \mathbf{1}$. Conversely, if we start with a Hermitian form $g_{\alpha\beta'}$ and a non-singular Hermitian matrix $\bm{\omega}$, then the tensor defined by
\begin{equation} \label{gTrans}
\tilde g_{\alpha\beta'} = \omega_\alpha^\beta g_{\beta\beta'}, \qquad \tilde g^{\alpha\beta'} = \Omega^\alpha_\beta g^{\beta\beta'}
\end{equation}
will also be a Hermitian form.

It may seem that a strange paradox arises here: on one hand, if we denote the matrix of the transformation $g_1\mapsto g_2$ as $\bm{\omega}_{21}$ (in this paragraph we will omit the indices), then it follows from the formula \eqref{gTrans} that for any three Hermitian forms $g_1$, $g_2$, and $g_3$ the relation holds
\begin{equation} \label{omegaComp}
\bm{\omega}_{31} = \bm{\omega}_{32} \cdot \bm{\omega}_{21}.
\end{equation}
However, on the other hand, it is well-known that the product of two Hermitian matrices is not necessarily Hermitian itself. This apparent contradiction is resolved very simply: the operation of Hermitian conjugation can be defined only with respect to some given Hermitian form. Therefore, when we say that some matrix is Hermitian, we must always specify: with respect to which Hermitian form? So the matrix $\bm{\omega}_{21}$ is Hermitian simultaneously with respect to $g_1$ and $g_2$, but not necessarily with respect to some third Hermitian form $g_3$. If we now return to the formula \eqref{omegaComp} above, we notice that whichever of the three Hermitian forms we take, two matrices in the formula will be Hermitian with respect to it, but the third will not. Thus, no contradiction arises.

In complete analogy with \eqref{InfinitesimalTransforms}-\eqref{TaylorSeries} we can write:
\begin{equation} \label{hDef}
\bm{\omega} = \exp(\bm{h}), \qquad  \bm{\Omega} = \exp(-\bm{h}),
\end{equation}
where $\bm{h} = \bm{h}^\dag$ is also a Hermitian matrix (both with respect to $g_{\alpha\beta'}$ and $\tilde g_{\alpha\beta'}$).

%%%%%%%%%%%%%%%%%%%%%%%%%%%%%%%%%%%%%%%%%%%
\subsection{General field transformations}

So, when we perform general field transformations
\begin{equation} \label{GeneralTransformation}
(\nabla_a, g_{\alpha\beta'}) \mapsto (\tilde\nabla_a, \tilde g_{\alpha\beta'}),
\end{equation}
we transform simultaneously the connection using a total potential $\calA_a[\tilde\nabla - \nabla]$ \eqref{RaznDiff} and the Hermitian form using a pair of mutually inverse Hermitian matrices $\bm{\omega}$ and $\bm{\Omega}$ \eqref{gTrans}.

It seems that we can, in complete analogy with \eqref{TotalCurvature}-\eqref{FGdef}, split the total potential $\calA_a$ into Hermitian and anti-Hermitian parts
\begin{align}
&\calA_a = \bm{B}_a - i\bm{A}_a, \quad\text{where} \label{TotalPotential} \\
&\bm{B}_a = \Hrm\calA_a, \qquad \bm{A}_a = \aHrm\calA_a. \label{ABdef}
\end{align}

However, here we are faced with the question: how should we split $\calA_a[\tilde\nabla-\nabla]$ into Hermitian and anti-Hermitian parts---using $g_{\alpha\beta'}$ or $\tilde g_{\alpha\beta'}$? Both ways are equivalent, and so for definiteness and consistency we must adopt some convention. Let us agree that in all cases without exception, when the Hermitian form is transformed \eqref{gTrans}, we will raise and lower indices and, accordingly, split into the Hermitian and anti-Hermitian parts only with the help of the old Hermitian form $g_{\alpha\beta'}$. Although this does not affect the result in the infinitesimal case, inattention in this regard can easily lead to errors when we make finite transformations. We would even recommend, in order to avoid them, to carry out all variations and other calculations in terms of the total potential $\calA_a$, and to pass to its components $\bm{A}_a$ and $\bm{B}_a$ only in the final expressions.

Let us now derive how the YM-deviation vector $\bm{N}_a[\nabla, g]$ is transformed under \eqref{GeneralTransformation}. From the definition \eqref{YMdeviationDef} and transformation laws \eqref{RaznDiff} and \eqref{gTrans}, we obtain
\begin{align}
&\bm{N}_a[\tilde\nabla, \tilde g] \cong -\frac{1}{2}\tilde g^{\beta\beta'} \tilde\nabla_a \tilde g_{\alpha\beta'} = \frac{1}{2}\Omega_\sigma^\beta g^{\sigma\beta'} \nonumber \\
&\times \big(-\nabla_a (\omega_\alpha^\delta g_{\delta\beta'}) + \calA_a{}_\alpha^\gamma \omega_\gamma^\delta g_{\delta\beta'} + \bar\calA_a{}_{\beta'}^{\gamma'} \omega_\alpha^\delta g_{\delta\gamma'}\big) \nonumber \\
&\cong \bm{\Omega} \bm{N}_a[\nabla, g] \bm{\omega} - \frac{1}{2} \bm{\Omega} \nabla_a \bm{\omega} + \frac{1}{2} \big(\calA_a + \bm{\Omega} \calA_a^\dag \bm{\omega}\big). \label{Ntransformation}
\end{align}
(Note that $\bm{N}_a[\tilde\nabla, \tilde g]$ will be Hermitian with respect to $\tilde g_{\alpha\beta'}$, but not with respect to $g_{\alpha\beta'}$.)

Going to the infinitesimal form, we obtain the following variations:
\begin{align}
\delta_h \bm{N}_a &= -\frac{1}{2}\nabla_a\bm{h} + [\bm{N}_a, \bm{h}], \label{Nvarg} \\
\delta_B \bm{N}_a &= \bm{B}_a, \qquad \delta_A\bm{N}_a = 0. \label{NvarAB}
\end{align}

Note that since the total curvature $\calF_{ab}[\nabla]$ does not depend on the Hermitian form at all, under the general transformation \eqref{GeneralTransformation} it will simply be transformed according to the formula \eqref{FPreobr}, which we rewrite here as
\begin{align}
&\calF_{ab}[\tilde\nabla] - \calF_{ab}[\nabla] = \calD_{ab} + \calC_{ab}, \label{FTrans2} \\
&\calD_{ab} = \nabla_a\calA_b - \nabla_b\calA_a, \qquad \calC_{ab} = [\calA_a, \calA_b]. \label{calCD}
\end{align}

To split $\calD_{ab}$ and $\calC_{ab}$ into Hermitian and anti-Hermitian parts, it is convenient to introduce the following combinations
\begin{align}
\check{\bm D}_{ab} &= \nabla_a\bm{B}_b - \nabla_b\bm{B}_a, \qquad \check{\bm C}_{ab} = i[\bm{B}_a, \bm{B}_b], \label{checkDC} \\
\hat{\bm D}_{ab} &= \nabla_a\bm{A}_b - \nabla_b\bm{A}_a,  \qquad \hat{\bm C}_{ab} = i[\bm{A}_a, \bm{A}_b], \label{hatDC} \\
\bm{C}_{ab} &= i[\bm{A}_a, \bm{B}_b] - i[\bm{A}_b, \bm{B}_a], \label{bmCdef} \\
\check{\bm K}_{ab} &= i[\bm{N}_a, \bm{B}_b] - i[\bm{N}_b, \bm{B}_a], \\
 \hat{\bm K}_{ab} &= i[\bm{N}_a, \bm{A}_b] - i[\bm{N}_b, \bm{A}_a].
\end{align}

Then we have the following decomposition of $\calC_{ab}$ 
\begin{align}
\check\calC_{ab} &= \Hrm\calC_{ab} = -\bm{C}_{ab}, \\
\hat\calC_{ab} &= \aHrm\calC_{ab} = \check{\bm C}_{ab} - \hat{\bm C}_{ab}.
\end{align}
For $\calD_{ab}$, the decomposition will be slightly more complicated (due to \eqref{DerHermEq}):
\begin{align}
\check\calD_{ab} &= \Hrm\calD_{ab} = \check{\bm D}_{ab} + i\calK_{ab}, \label{DeltaF1} \\
\hat\calD_{ab} &= \aHrm\calD_{ab} = \hat{\bm D}_{ab} + \calK_{ab}, \label{DeltaF2} \\
\calK_{ab} &= \check{\bm K}_{ab} + i \hat{\bm K}_{ab}.
\end{align}
These relations have the following meaning: the matrices $\check{\bm D}_{ab}$ and $\hat{\bm D}_{ab}$ are not Hermitian, and, accordingly, the commutators with $\bm{N}_a$ in $\calK_{ab}$ restore the hermiticity of the entire expressions. Let us emphasize once again that, in accordance with the agreement at the beginning of this subsection, here we perform all partitions with respect to the same old Hermitian form $g_{\alpha\beta'}$, not to the new $\tilde g_{\alpha\beta'}$.

The Hermitian conjugate of the total curvature $\calF^\dag_{ab}[\nabla, g]$ depends on both $\nabla_a$ and $g_{\alpha\beta'}$. Its transformation under a change of the Hermitian form can be written as
\begin{equation} \label{calFdagTrans}
\calF_{ab}^\dag[\tilde\nabla, \tilde g] = \bm{\Omega}\calF_{ab}^\dag[\tilde\nabla, g] \bm{\omega}.
\end{equation}

Accordingly, to obtain transformations of the Hermitian and anti-Hermitian components of curvature $\bm{G}_{ab}$ and $\bm{F}_{ab}$, we must act according to the following algorithm:
\begin{enumerate}
\item Using formulas \eqref{TotalCurvature}-\eqref{FGdef}, assemble from them the complex quantities $\calF_{ab}$ and $\calF_{ab}^\dag$ (here, the ``old'' Hermitian form $g_{\alpha\beta'}$ is implied);

\item Change the connection $\nabla_a\mapsto\tilde\nabla_a$, transforming $\calF_{ab}$ according to \eqref{FTrans2} (and $\calF_{ab}^\dag$, respectively, using the Hermitian conjugate relation);

\item Change the Hermitian form $g_{\alpha\beta'}\mapsto \tilde g_{\alpha\beta'}$, transforming $\calF_{ab}^\dag$ by the formula \eqref{calFdagTrans};

\item Finally, apply formulas \eqref{TotalCurvature}-\eqref{FGdef} in the opposite direction to obtain the transformed $\bm{G}_{ab}$ and $\bm{F}_{ab}$ (here, the ``new'' Hermitian form $\tilde g_{\alpha\beta'}$ is implied).
\end{enumerate}

The result of this algorithm can be represented as a combination of two steps. In the first one, we transform only the connection, introducing the following auxiliary quantities:
\begin{align}
\bm{G}'_{ab} &= \bm{G}_{ab}[\nabla, g] + \check{\bm D}_{ab} - \bm{C}_{ab}, \label{FinalFGtrans1} \\
\bm{F}'_{ab} &= \bm{F}_{ab}[\nabla, g] + \hat{\bm D}_{ab} + \check{\bm C}_{ab} - \hat{\bm C}_{ab}; \label{FinalFGtrans2}
\end{align}
and in the second, we change the Hermitian form
\begin{align}
\bm{G}_{ab}[\tilde\nabla, \tilde g] &= \frac{1}{2}\big(\bm{G}'_{ab} + \bm{\Omega}\bm{G}'_{ab}\bm{\omega}\big) \label{FinalFGtrans3} \\
&-\frac{i}{2}\big(\bm{F}'_{ab} - \bm{\Omega}\bm{F}'_{ab}\bm{\omega}\big) + i\bm{\Omega}\calK_{ab}\bm{\omega}, \nonumber \\
\bm{F}_{ab}[\tilde\nabla, \tilde g] &= \frac{1}{2}\big(\bm{F}'_{ab} + \bm{\Omega}\bm{F}'_{ab}\bm{\omega}\big) \label{FinalFGtrans4} \\
&+\frac{i}{2}\big(\bm{G}'_{ab} - \bm{\Omega}\bm{G}'_{ab}\bm{\omega}\big) + \bm{\Omega}\calK_{ab}\bm{\omega}, \nonumber
\end{align}

If we do not transform the Hermitian form, $\bm{\Omega} = \bm{\omega} = \bm{1}$, these expressions simplify significantly
\begin{equation}
\bm{G}_{ab}[\tilde\nabla, \tilde g] = \bm{G}'_{ab} + i\calK_{ab}, \quad \bm{F}_{ab}[\tilde\nabla, \tilde g] = \bm{F}'_{ab} + \calK_{ab}.
\end{equation}
Discarding the terms from the quadratic part $\calC_{ab}$ in this expression, we obtain the following variations
\begin{align}
\delta_A \bm{F}_{ab} &= \hat{\bm D}_{ab} + i\hat{\bm K}_{ab}, & \delta_A \bm{G}_{ab} &= -\hat{\bm K}_{ab}, \label{GvarA}\\
\delta_B \bm{G}_{ab} &= \check{\bm D}_{ab} + i\check{\bm K}_{ab}, & \delta_B \bm{F}_{ab} &= \check{\bm K}_{ab}. \label{GvarB}
\end{align}
As we can see, the presence of a non-zero YM-deviation vector $\bm{N}_a \ne 0$ leads to a peculiar mixing of the Hermitian and anti-Hermitian components: $\bm{A}_a$ begins to contribute to $\bm{G}_{ab}$, and $\bm{B}_a$---to $\bm{F}_{ab}$.

Variation with respect to $\bm{h}$ can easily be derived from the infinitesimal form of transformation \eqref{calFdagTrans}
\begin{equation}
\delta_h \calF_{ab}^\dag = [\calF_{ab}^\dag, \bm{h}].
\end{equation}
Splitting it into Hermitian and anti-Hermitian parts, we obtain
\begin{align}
\delta_h \bm{G}_{ab} &= \frac{1}{2}[\bm{G}_{ab}, \bm{h}] + \frac{i}{2}[\bm{F}_{ab}, \bm{h}], \label{Gvarg} \\
\delta_h \bm{F}_{ab} &= \frac{1}{2}[\bm{F}_{ab}, \bm{h}] - \frac{i}{2}[\bm{G}_{ab}, \bm{h}]. \label{Fvarg}
\end{align}

One can check that infinitesimal transformations \eqref{Nvarg}-\eqref{NvarAB}, \eqref{GvarA}-\eqref{GvarB} and \eqref{Gvarg}-\eqref{Fvarg} are indeed consistent with the key relation \eqref{NtoG}.

%%%%%%%%%%%%%%%%%%%%%%%%%%%%%%%%%%%%%%%%%%%
\section{Gauge symmetries and Noether identities} \label{GaugeSec}
%%%%%%%%%%%%%%%%%%%%%%%%%%%%%%%%%%%%%%%%%%%

%%%%%%%%%%%%%%%%%%%%%%%%%%%%%%%%%%%%%%%%%%%%%%%
\subsection{$\mathrm{GL}(n,\bbC) \to \mathrm{U}(n)$ gauge symmetry breaking} \label{GaugeSymBreaking}

In Subsec.~\ref{InvTransformSect} we considered the general $\mathrm{GL}(n,\bbC)$ gauge symmetry of any theory to consist of invertible linear transformations in the internal color space \eqref{InvColorTransform} together with the corresponding connection transformation \eqref{PureGaugePotencalEq}. We will now consider what happens if a Hermitian form $g_{\alpha\beta'}$ is additionally specified.

First of all, we note that these transformations, generally speaking, change the Hermitian form:
\begin{equation}
g_{\alpha\alpha'} \mapsto \tilde g_{\alpha\alpha'} = u_\alpha^\beta \bar u_{\alpha'}^{\beta'} g_{\beta\beta'}.
\end{equation}
Substituting this expression into \eqref{OmegaMatrices}, we immediately obtain the Hermitian form transformation matrices
\begin{equation} \label{UtoOmega}
\bm{\omega} = \bm{u}^\dag \bm{u}, \qquad \bm{\Omega} = \bm{U} \bm{U}^\dag.
\end{equation}
Therefore, the Hermitian form remains unchanged ($\bm{\omega} = \bm{\Omega} = \mathbf{1}$) if the condition holds
\begin{equation} \label{UnitaryCondition}
\bm{U} = \bm{u}^\dag,
\end{equation}
i.e. when the transformations of the internal color space are \emph{unitary}.

If we now consider infinitesimal transformations and split the transformation parameter $\bm{\epsilon}$ \eqref{InfinitesimalTransforms} into anti-Hermitian $\bm{\alpha}$ and Hermitian $\bm{\beta}$ parts
\begin{equation} \label{aHrmGaugeTrans}
\bm{\epsilon} = \bm{\beta} - i\bm{\alpha}, \qquad \bm{\beta} = \Hrm\bm{\epsilon}, \qquad \bm{\alpha} = \aHrm\bm{\epsilon},
\end{equation}
then it is easy to see that the Hermitian form perturbation \eqref{hDef} looks like
\begin{equation} \label{GaugeH}
\bm{h} = 2\bm{\beta}.
\end{equation}
So unitary transformations correspond to the anti-Hermitian part $\bm{\alpha}$, and non-unitary transformations correspond to the Hermitian part $\bm{\beta}$.

Thus, we conclude that \emph{the Hermitian form $g_{\alpha\beta'}$ in mal-YM acts as a Higgs field}: any theory has the general $\mathrm{GL}(n, \bbC)$ gauge symmetry given by the transformations \eqref{InvColorTransform} and \eqref{PureGaugePotencalEq}. However, introducing the Hermitian form $g_{\alpha\beta'}$ breaks this symmetry to the group $U(n)$ of unitary transformations, which keep $g_{\alpha\beta'}$ unchanged.

Now, if we take the expression for the potential corresponding to the full gauge transformation $\calA_a[\tilde\nabla - \nabla] = \nabla_a\bm{\epsilon}$ \eqref{InfinitesimalTransforms} and expand it into Hermitian and anti-Hermitian parts (using the formula \eqref{DerHermEq}), we get:
\begin{align}
\bm{A}_a &= \nabla_a\bm{\alpha} - [\bm{N}_a, \bm{\alpha}] + i[\bm{N}_a, \bm{\beta}], \label{GaugeA} \\
\bm{B}_a &= \nabla_a\bm{\beta} - [\bm{N}_a, \bm{\beta}] - i[\bm{N}_a, \bm{\alpha}]. \label{GaugeB}
\end{align}
In this case, tensors $\bm{N}_a$, $\bm{F}_{ab}$ and $\bm{G}_{ab}$ will, like all matrices, be transformed according to the general rule \eqref{MatrixTrans}:
\begin{equation} \label{TensGaugeTrans}
\delta\bm{N}_a = [\bm{N}_a, \bm{\epsilon}], \quad \delta\bm{F}_{ab} = [\bm{F}_{ab}, \bm{\epsilon}], \quad \delta\bm{G}_{ab} = [\bm{G}_{ab}, \bm{\epsilon}].
\end{equation}

It is clear that these transformations leave any Lagrangian built from tensors $\bm{N}_a$, $\bm{F}_{ab}$ and $\bm{G}_{ab}$ invariant. Therefore they are genuine (infinitesimal) gauge transformations of any theory.

(One can directly obtain the transformations \eqref{TensGaugeTrans} by substituting the expressions \eqref{GaugeH}-\eqref{GaugeB} into the variations \eqref{Nvarg}-\eqref{NvarAB}, \eqref{GvarA}-\eqref{GvarB} and \eqref{Gvarg}-\eqref{Fvarg} and using the relation \eqref{NtoG} and the Jacobi identity for commutators.)

It is curious to note that although, as we have noted above, the general transformations of fields \eqref{GvarA}-\eqref{GvarB} and \eqref{Gvarg}-\eqref{Fvarg} mix the Hermitian and anti-Hermitian components, this mixing is absent from the gauge transformation law in a sense that the fields $\bm{F}_{ab}$ and $\bm{G}_{ab}$ transform independently of each other. On the other hand, it may seem that these transformations \eqref{TensGaugeTrans} are inconsistent. Indeed, for non-unitary transformations ($\bm{\alpha} = 0$, $\bm{\epsilon} = \bm{\beta}$) it follows that the variation of the Hermitian matrix is anti-Hermitian. But there is no error or contradiction here. This phenomenon is simply due to the fact, also noted above, that gauge transformations with a Hermitian parameter $\bm{\beta}$ change the Hermitian form $g_{\alpha\beta'}$ and, consequently, the very concept of hermiticity. For example, the transformed tensor $\bm{G}_{ab}[\tilde\nabla, \tilde g]$ will, of course, be a Hermitian component of the total curvature with respect to the new Hermitian form $\tilde g_{\alpha\beta'}$, but not with respect to the original $g_{\alpha\beta'}$; accordingly, the variation $\delta\bm{G}_{ab} = \bm{G}_{ab}[\tilde\nabla, \tilde g] - \bm{G}_{ab}[\nabla, g]$ will not be such either.

We will continue the discussion of gauge transformations below in Subsections \ref{NoetherSubsec} and \ref{EatingGoldstones}.

%%%%%%%%%%%%%%%%%%%%%%%%%%%%%%%%%%%%%%%%%%%%%%%%%
\subsection{Background field method} \label{BackgroundSubsec}

Before we move on to our specific model, a brief reminder of the background field method that we will use in what follows is necessary. Let a theory be given with a set of fields $\varphi = \varphi^A$ whose dynamics are determined by some given classical action functional
\begin{equation} \label{ActionEq}
S[\varphi] = \int d^dx\, \calL(\varphi, \partial\varphi,\dots).
\end{equation}

The idea is to represent the fields as
\begin{equation}
\varphi = \Phi + \phi,
\end{equation}
where $\Phi$ are fixed \emph{background fields} and $\phi$ are their small \emph{perturbations}. So $S[\varphi]$ becomes a functional of both $\Phi$ and $\phi$, and we expand it into a functional Taylor series in perturbations~$\phi$:
\begin{align}
&S[\Phi + \phi] = \sum\limits_{n=0}^\infty \frac{1}{n!} S_n[\Phi] \cdot \phi^n, \quad\text{where} \label{PerturbationSeries} \\
&S_n[\Phi] \cdot \phi^n = \int dx_1\cdots dx_n\,S_n[\Phi|x_1,\ldots, x_n] \nonumber \\
&\qquad\qquad\qquad\times\phi(x_1)\cdots \phi(x_n), \\
&S_n[\Phi|x_1,\ldots, x_n] = \left.\frac{\delta^n S[\varphi]}{\delta\varphi(x_1)\cdots\delta\varphi(x_n)}\right|_{\varphi = \Phi}.
\end{align}

Then the \emph{classical $n$-point functions} $S_n[\Phi]$ are functionals of the background fields $\Phi$. They have an extremely simple physical meaning:
\begin{itemize}
\item $S_0[\Phi]$ is just the original classical action for background fields;
\item $S_1[\Phi]$ contains information about the sources of the fields, or, in other words, defines classical equations of motion (EoMs) for background fields;
\item $S_2[\Phi]$ gives the linearized EoMs for infinitesimal perturbations $\phi$ on a given background $\Phi$, or, in other words, determines the propagators of the theory;
\item Finally, $S_n[\Phi]$ for $n\ge3$ describe the nonlinear interaction of perturbations $\phi$ or, what is the same, the $n$-vertices of the theory.
\end{itemize}
Both the propagators and the vertices, generally speaking, depend on background fields $\Phi$.

Now let us return to our consideration of mal-YM. In this particular case, the partition into background fields and perturbations will take the following form:
\begin{itemize}
\item \emph{background fields}: this is the Hermitian form $g_{\alpha\beta'}$ and the connection $\nabla_a$---and also all functions of them, such as the total curvature in the bundle $\calF_{ab}[\nabla]$ \eqref{DefF}, its Hermitian $\bm{G}_{ab}[\nabla, g]$ and anti-Hermitian $\bm{F}_{ab}[\nabla, g]$ parts \eqref{TotalCurvature}-\eqref{FGdef}, and the YM-deviation vector $\bm{N}_a[\nabla, g]$ \eqref{YMdeviationDef};

\item \emph{perturbations:} this is the change of the Hermitian form $\bm{h}$ \eqref{hDef} and the total potential $\calA_a$ \eqref{GaugePotencialDef}---and also its Hermitian $\bm{B}_a$ and the anti-Hermitian $\bm {A}_a$ parts \eqref{TotalPotential}-\eqref{ABdef}.
\end{itemize}

Then, applying the background field method to the action of our theory $S[\nabla, g]$ we successively find the classical EoMs, the propagators, and the interaction vertices. It turns out that it is much more convenient to initially write the action in terms of the total curvature $\calF_{ab}$ and vary it over the total potential $\calA_a$. And only at the very last step rewrite the obtained answer in terms of their Hermitian and anti-Hermitian parts $\bm{F}_{ab}$, $\bm{G}_{ab}$, $\bm{A}_a$, and~$\bm{B}_a$.

%%%%%%%%%%%%%%%%%%%%%%%%%%%%%%%%%%%%%%%%%%%%
\subsection{Field sources and Noether identities} \label{NoetherSubsec}

Given some action $S[\Phi]$, the source of the background field $\Phi^A$ is defined as the variational derivative $J_A[\Phi] = \delta S[\Phi+\phi]/\delta\phi^A |_{\phi = 0}$. For example, the total potential $\calA_a$ will be associated with \emph{the total gauge current}
\begin{equation} \label{totalCurrent}
\calJ^a = -2\frac{\delta S}{\delta\calA_a}.
\end{equation}

Splitting it into Hermitian and anti-Hermitian parts in the usual way, we get
\begin{equation} \label{aHrmCurrents}
\bm{\Lambda}^a = \Hrm\calJ^a = -\frac{\delta S}{\delta\bm{B}_a}, \quad \bm{J}^a = \aHrm\calJ^a = \frac{\delta S}{\delta\bm{A}_a}.
\end{equation}
The appearance of a new Hermitian gauge current $\bm{\Lambda}^a$ and the corresponding new EoM naturally follows from the appearance of the Hermitian component of the potential $\bm{B}_a$ and is an essential new feature of mal-YM.

In addition, we can define a variation of the action with respect to the Hermitian form perturbation
\begin{equation} \label{hSource}
\bm{E} = -2 \frac{\delta S}{\delta\bm{h}},
\end{equation}
which is a Hermitian matrix $\bm{E}^\dag = \bm{E}$. To the same extent that the Hermitian form $g_{\alpha\beta'}$ is an analogue of the spacetime metric $g_{ab}$, this matrix $\bm{E}$ is similar in its properties to the stress-energy $T_{ab}$ and the Einstein $G_{ab}$ tensors  obtained by varying the action with respect to the metric.

However, it is important to note the following: it is well known that if a theory has a gauge symmetry, then it gives rise to \emph{Noether identities}---special relations between sources of fields.

Consider the case of pure mal-YM, when there are no additional matter fields. Then, under infinitesimal $GL(n,\mathbb{C})$ gauge transformations with parameter $\bm{\epsilon}$ \eqref{InfinitesimalTransforms} the total potential $\calA_a$ and the variation of the Hermitian form are given by \eqref{InfTensTrans} and~\eqref{GaugeH} respectively. By construction, these transformations cannot change any geometrically-defined action. Then for the variation of the action we will have
\begin{align}
\delta_\epsilon S[\nabla, g] &= \int d^dx \tr\left(\frac{\delta S}{\delta\calA_a} \calA_a + \frac{\delta S}{\delta\calA_a^\dag} \calA_a^\dag + \frac{\delta S}{\delta\bm{h}} \bm{h} \right) \nonumber \\
&= -\frac{1}{2} \int d^dx \tr\left(\calJ_a \nabla^a \bm{\epsilon} + \calJ_a^\dag (\nabla^a\bm{\epsilon})^\dag + 2\bm{E}\bm{\beta} \right) \nonumber \\
&= -\frac{1}{2} \int d^dx \tr\Big((\bm{E} - \nabla^a\calJ_a)\bm{\epsilon} \nonumber \\
&+ (\bm{E} - \nabla^a\calJ_a^\dag + 2[\bm{N}^a, \calJ_a^\dag]) \bm{\epsilon}^\dag \Big) = 0.
\end{align}
Since the gauge parameter $\bm{\epsilon}$ here is an arbitrary complex matrix, we obtain the following Noether identity: the Hermitian matrix $\bm{E}$ is equal to the divergence of the total gauge current $\calJ_a$
\begin{equation}
\bm{E} = \nabla^a \calJ_a = \nabla^a \calJ_a^\dag - 2[\bm{N}^a, \calJ_a^\dag]. \label{NoetherId}
\end{equation}
Or if we split these relations into anti-Hermitian and Hermitian parts:
\begin{align}
&\nabla_a\bm{J}^a - [\bm{N}_a, \bm{J}^a] + i[\bm{N}_a, \bm{\Lambda}^a] = 0, \\
&\nabla_a\bm{\Lambda}^a - [\bm{N}_a, \bm{\Lambda}^a] - i[\bm{N}_a, \bm{J}^a] = \bm{E}.
\end{align}
The first of these relations is a modification of the conservation law of the standard YM gauge current, which curiously ceases to be preserved in mal-YM. And the second relation means that the equation $\bm{E} = 0$ obtained by varying the action with respect to $\bm{h}$ is not an independent EoM, but only a differential consequence of $\bm{J}_a = \bm{\Lambda}_a = 0$.

%%%%%%%%%%%%%%%%%%%%%%%%%%%%%%%%%%%%%%%%%%%
\section{Action} \label{ActionSec}
%%%%%%%%%%%%%%%%%%%%%%%%%%%%%%%%%%%%%%%%%%%

%%%%%%%%%%%%%%%%%%%%%%%%%%%%%%%%%%%%%%%%%%%%%%%%%
\subsection{Mal-YM action}

Now we concretize the action of the theory that we will consider. From only curvature tensor $\calF_{ab}$, we can construct the following real scalar combination of dimension~4\footnote{We use units of mass dimension $[\nabla_a] = 1$, $[g_{\alpha\beta'}] = 0$. Then we have $[\bm{F}_{ab}] = [\bm{G}_{ab}] = 2$, $[\bm{N}_a] = 1$. The unusual sign here is due to the fact that we assume Euclidean field theory throughout; the transition to the physical Lorentz signature is accomplished via the Wick rotation.}:
\begin{equation} \label{Lagr1}
\calL_1[\nabla] = \frac{1}{8}\tr\left(\calF_{ab}\calF^{ab} + \bar\calF_{ab}\bar\calF^{ab}\right).
\end{equation}
This combination depends only on the connection $\nabla_a$, but not on the Hermitian form $g_{\alpha\beta'}$. If we involve $g_{\alpha\beta'}$, then we can construct another combination of dimension 4 that will depend on both quantities:
\begin{equation} \label{Lagr2}
\calL_2[\nabla, g] = \frac{1}{4}\tr(\calF_{ab}^\dag \calF^{ab}).
\end{equation}
In addition, we also want to add some term that would allow us to restore the standard YM in the limit. The most natural way that first comes to our mind is to try to use for this purpose the YM-deviation vector squared
\begin{equation} \label{Lag3}
\calL_3[\nabla, g] = \calL_{N^2} = \frac{1}{2}\tr\big(\bm{N}_a \bm{N}^a\big).
\end{equation}

Of course, we could also introduce other terms, for example, $i\tr(\calF_{ab}\calF^{ab} - \bar\calF_{ab}\bar\calF^{ab})/4 = \tr(\bm{G}_{ab}\bm{F}^{ab})$, $\tr([\bm{N}_a, \bm{N}_b]\bm{F}^{ab})$, etc., but we do not do this solely for the sake of simplicity, in order not to clutter the presentation.
Thus we arrive at the following total Lagrangian, which we will use in what follows:
\begin{equation} \label{malYMlagrangian}
\calL_\text{mal-YM} = c_1 \calL_1 + c_2 \calL_2 + c_3 \calL_3.
\end{equation}

If we rewrite this expression in terms of $\bm{G}_{ab}$ and $\bm{F}_{ab}$, we obtain the following form of the total Lagrangian
\begin{equation} \label{totalLagrangian}
\calL_\text{mal-YM} = \frac{1}{e^2}\calL_{F^2} + \frac{1}{\tilde e^2} \big(\calL_{G^2} + M^2 \calL_{N^2}\big),
\end{equation}
where
\begin{gather}
\calL_{F^2} = \frac{1}{4} \tr\big(\bm{F}_{ab} \bm{F}^{ab}\big) = \frac{1}{2}(\calL_2 - \calL_1), \label{LagF} \\
\calL_{G^2} = \frac{1}{4} \tr\big(\bm{G}_{ab} \bm{G}^{ab}\big) = \frac{1}{2}(\calL_2 + \calL_1), \label{LagG} \\
\frac{1}{e^2} = c_2 - c_1, \quad \frac{1}{\tilde e^2} = c_1 + c_2, \quad M^2 = \frac{c_3}{c_1+c_2}.
\end{gather}

Since the action \eqref{ActionEq} must be dimensionless, in $d$-dimensional spacetime we have $[\calL] = d$, which implies the following dimensions of the coupling constants of our theory $[c_1] = [c_2] = d-4$, $[c_3] = d-2$. In the usual four-dimensional case $d=4$, the constants $e$ and $\tilde e$ become dimensionless, and $M$ acquires the dimension 1.

Note that we can vary the three terms in \eqref{malYMlagrangian} separately, and only then combine the results obtained in this way. Therefore, we will use the following notation: the currents generated by the corresponding term in the action will be denoted, as above, by a subscript; and the Lagrangian with a superscript in parentheses will denote the term in the expansion \eqref{PerturbationSeries} with the corresponding power in perturbations.

%%%%%%%%%%%%%%%%%%%%%%%%%%%%%%%%%%%%%%%%%%%%
\subsection{EoMs for background fields}

Let us start with classical field sources and the corresponding EoMs for background fields. For each term in the action, we write out the linear part in perturbations, where the antisymmetric combination $\calD_{ab}$ and its Hermitian $\check{\calD}_{ab}$ and anti-Hermitian $\hat{\calD}_{ab}$ parts were defined above \eqref{calCD}, \eqref{DeltaF1}-\eqref{DeltaF2}.

For the first term \eqref{Lagr1}, independent of $g_{\alpha\beta'}$, we have
\begin{align}
\calL_1^{(1)} &= \frac{1}{4} \tr\left(\calF_{ab}\calD^{ab} + \bar{\calF}_{ab} \bar{\calD}^{ab}\right) \nonumber \\
&= \frac{1}{2} \tr\left(\bm{G}_{ab}\check{\calD}^{ab} - \bm{F}_{ab}\hat{\calD}^{ab}\right),
\end{align}
from which we obtain the following expressions for the sources:
\begin{equation} \label{Sources1}
\calJ^a_1 = \nabla_b\calF^{ba}, \qquad \bm{E}_1 = 0.
\end{equation}

For the second term \eqref{Lagr2} there will be an additional term, linear in $\bm{h}$
\begin{align}
&\calL_h^{(1)} = \frac{1}{4} \tr\left( [\calF^{ab}, \calF_{ab}^\dag] \bm{h}\right) = \frac{i}{2}\tr\left([\bm{G}_{ab}, \bm{F}^{ab}] \bm{h}\right), \\
&\calL_2^{(1)} - \calL_h^{(1)} = \frac{1}{4} \tr\left(\calF^{ab} \calD_{ab}^\dag + \calF_{ab}^\dag\calD^{ab}\right) \nonumber \\
&\qquad\qquad\;\;\;= \frac{1}{2} \tr\left(\bm{G}_{ab}\check{\calD}^{ab} + \bm{F}_{ab}\hat{\calD}^{ab}\right).
\end{align}
These terms will generate sources
\begin{equation} \label{Sources2}
\calJ^a_2 = \nabla_b\calF^{ba\dag}, \quad \bm{E}_2 = \frac{1}{2}[\calF_{ab}^\dag, \calF^{ab}] = i [\bm{F}_{ab}, \bm{G}^{ab}].
\end{equation}

Accordingly, for combinations \eqref{LagF}-\eqref{LagG} we have
\begin{align}
\calL^{(1)}_{G^2} &= \frac{1}{2}\tr\big(\bm{G}_{ab}\check{\calD}^{ab}\big) + \frac{1}{2}\calL_h^{(1)}, \\
\calL^{(1)}_{F^2} &= \frac{1}{2}\tr\big(\bm{F}_{ab}\hat{\calD}^{ab}\big)  + \frac{1}{2}\calL_h^{(1)},
\end{align}
and for the Hermitian and anti-Hermitian components of the gauge current generated by them we will have
\begin{align}
\bm{\Lambda}^a_{G^2} &= \nabla_b\bm{G}^{ba} - [\bm{N}_b, \bm{G}^{ba}], & \bm{\Lambda}^a_{F^2} &= i[\bm{N}_b, \bm{F}^{ba}], \\
\bm{J}^a_{F^2} &= \nabla_b\bm{F}^{ab} + [\bm{N}_b, \bm{F}^{ba}], & \bm{J}^a_{G^2} &= i[\bm{N}_b, \bm{G}^{ba}].
\end{align}

Finally, for the third term \eqref{Lag3} we have
\begin{gather}
\calL_3^{(1)} = \tr\Big((\bm{B}_a - \frac{1}{2} \nabla_a\bm{h}) \bm{N}^a\Big), \\
\calJ^a_3 = \bm{\Lambda}^a_3 = -\bm{N}^a, \quad \bm{J}^a_3 = 0, \quad \bm{E}_3 = -\nabla_a\bm{N}^a. \label{Sources3}
\end{gather}
It is immediately evident that for each of the three terms the sources \eqref{Sources1}, \eqref{Sources2}, and \eqref{Sources3} generated by them satisfy the Noether identities \eqref{NoetherId}.

Combining these three terms yields the following EoM
\begin{equation} \label{EquationOfMotion}
c_1 \nabla_b\calF^{ab} + c_2 \nabla_b\calF^{ab\dag} + c_3\bm{N}^a = \calJ^a_\text{ext}.
\end{equation}
Or if we split it into Hermitian and anti-Hermitian parts:
\begin{align}
&\nabla_b \bm{F}^{ab} - [\bm{N}_b, \bm{F}^{ab}] - i\frac{e^2}{\tilde e^2} [\bm{N}_b, \bm{G}^{ab}] = -e^2 \bm{J}^a_\text{ext}, \label{EoMF} \\
&\nabla_b \bm{G}^{ab} - [\bm{N}_b, \bm{G}^{ab}] + i\frac{\tilde e^2}{e^2} [\bm{N}_b, \bm{F}^{ab}] \nonumber \\
&\qquad\qquad\qquad\qquad\qquad\quad+ M^2\bm{N}^a = \tilde e^2 \bm{\Lambda}^a_\text{ext}. \label{EoMG}
\end{align}
Here on the right-hand side we have additionally added an external gauge current $\calJ^a_\text{ext} = \bm{\Lambda}^a_\text{ext} - i\bm{J}^a_\text{ext}$, which can be created, for example, by gauge-charged matter fields, if they are present in the theory (see Subsection \ref{ScalarSubsec}).

The third equation, obtained by varying the total action with respect to $\bm{h}$, is of the form
\begin{equation} \label{EoME}
c_3 \nabla_a \bm{N}^a + ic_2 [\bm{G}_{ab}, \bm{F}^{ab}] = \bm{E}_\text{ext}.
\end{equation}
However, as is clear from the discussion of Noether identities and can be easily verified directly, it is not an independent EoM, but only a differential consequence of \eqref{EoMF}-\eqref{EoMG}.

%%%%%%%%%%%%%%%%%%%%%%%%%%%%%%%%%%%%%%%%%%%%%%%%%%%
\subsection{Linearized EoMs for perturbations on the trivial background} \label{LinearizedEoMsSubsec}

On an arbitrary given background $\bm{F}_{ab}$ and $\bm{N}_a$ [and hence $\bm{G}_{ab}$ by \eqref{NtoG}], the expressions for the quadratic in perturbations part of the action and the linearized EoMs become quite cumbersome and not very instructive. Therefore, we present them separately in the Appendix \ref{PropApp}, and here we restrict ourselves to considering the trivial background $\bm{N}_a = 0$, $\bm{G}_{ab} = \bm{F}_{ab} = 0$.

In this case, the Hermitian form $g_{\alpha\beta'}$ is covariantly constant and covariant derivatives with respect to background connection commute---therefore, we write them as $\partial_a$ to emphasize it. Then the expressions \eqref{DeltaF1}-\eqref{DeltaF2} are significantly simplified
\begin{equation} \label{FreeQuadricPart}
\hat{\calD}_{ab} = \partial_a\bm{A}_b - \partial_b\bm{A}_a, \quad \check{\calD}_{ab} = \partial_a\bm{B}_b - \partial_b\bm{B}_a.
\end{equation}

Accordingly, the expressions for the quadratic in perturbations part of the Lagrangian take a particularly simple form:
\begin{align}
&\calL_{F^2}^{(2)} = \frac{1}{4}\tr(\hat{\calD}_{ab}\hat{\calD}^{ab}), \quad \calL_{G^2}^{(2)} = \frac{1}{4}\tr(\check{\calD}_{ab}\check{\calD}^{ab}), \label{TrivLinAction} \\
&\calL_{N^2}^{(2)} = \frac{1}{2}\tr\Big((\bm{B}_a - \frac{1}{2}\partial_a\bm{h})(\bm{B}^a - \frac{1}{2}\partial^a\bm{h})\Big). \label{TrivLinAction2}
\end{align}

Then the variation of the total Lagrangian $\calL_\text{mal-YM}^{(2)}$ with respect to the perturbations $\bm{A}_a$, $\bm{B}_a$, and $\bm{h}$ leads to the following system of equations:
\begin{align}
&(\delta_a^b \Box + \partial^b\partial_a) \bm{A}_b = 0, \label{LinEqA} \\
&(\delta_a^b \Box + \partial^b\partial_a) \bm{B}_b + M^2(\bm{B}_a - \frac{1}{2}\partial_a\bm{h}) = 0, \label{LinEqB} \\
&\Box\bm{h} + 2\partial_a\bm{B}^a = 0, \label{LinEqH}
\end{align}
where $\Box = -\partial_a\partial^a$ is the flat Laplacian.

Note that gauge invariance implies that, first, substituting the gauge transformations $\bm{h} = 2\bm{\beta}$ \eqref{GaugeH}, $\bm{A}_a = \partial_a\bm{\alpha}$ \eqref{GaugeA} and $\bm{B}_a = \partial_a\bm{\beta}$ \eqref{GaugeB} obviously turns these equations into identities. Second, the corresponding differential operators are degenerate, which necessitates a gauge fixing procedure, which we will consider below in Subsections \ref{EatingGoldstones} and \ref{xiGauge}. Finally, for $M^2\ne0$ the third equation \eqref{LinEqH} is not independent, but it is a differential consequence of the equation \eqref{LinEqB}.

%%%%%%%%%%%%%%%%%%%%%%%%%%%%%%%%%%%%%%%%%%%%%%%%%%%
\subsection{Analysis of the EoMs}

Let us look again at the exact nonlinear EoMs for the background fields \eqref{EoMF}-\eqref{EoME} and at the linearized equations for small perturbations on the trivial background \eqref{LinEqA}-\eqref{LinEqH}. In them, all fields are split into two (non-trivially interacting) sectors: the Yang-Mills $A$-sector, containing the usual YM fields $\bm{A}_a$ and $\bm{F}_{ab}$, and the St\"{u}ckelberg $B$-sector, containing the new fields $\bm{B}_a$, $\bm{h}$, $\bm{N}_a$, and $\bm{G}_{ab}$.

First of all, note that if we want to treat both fields $\bm{A}_a$ and $\bm{B}_a$ as dynamic, we need both terms in the action---$\calL_{F^2}$ and $\calL_{G^2}$. Indeed, if we have a single term $\calL_{F^2}$, then the field $\bm{G}_{ab}$ is not dynamic, obeying only the condition
\begin{equation}
[\bm{N}_b, \bm{F}^{ab}] = -ie^2 \bm{\Lambda}^a_\text{ext},
\end{equation}
and the field $\bm{F}_{ab}$ obeys the equation
\begin{equation}
\nabla_b \bm{F}^{ab} = -ie^2 \calJ^a_\text{ext},
\end{equation}
which differs from the usual YM equation only by the change in external sources of the field, to which the Hermitian gauge current $\bm{\Lambda}^a_\text{ext}$ will now also contribute. Conversely, if we keep only the term $\calL_{G^2}$, the situation becomes almost the opposite: now there are no restrictions on the field $\bm{F}_{ab}$ at all, and the field $\bm{G}_{ab}$ will obey the ``Yang-Mills-like'' equation $\nabla_b \bm{G}^{ab} = \tilde e^2 \calJ^a_\text{ext}$ with the additional condition $[\bm{N}_b, \bm{G}^{ab}] = -i\tilde e^2 \bm{J}^a_\text{ext}$.

However, if we include both terms---$\calL_{F^2}$ as well as $\calL_{G^2}$, then such a theory will describe \emph{two} non-trivially interacting gauge fields. In this case, as can be seen from the equations \eqref{LinEqA}-\eqref{LinEqB}, without the mass term $\calL_{N^2}$ both fields $\bm{A}_a$ and $\bm{B}_a$ will be massless, but after adding $\calL_{N^2}$ the field $\bm{B}_a$ will acquire mass $M$.

Now consider pure mal-YM (without any external sources $\calJ^a_\text{ext} = 0$). If we look for its solutions for which $\bm{N}_a = 0$ (and hence $\bm{G}_{ab} = 0$), then the remaining field $\bm{F}_{ab}$ will obey the usual YM equation $\nabla^b\bm{F}_{ab} = 0$. This means that \emph{every classical solution of pure YM is also a solution of pure mal-YM}. However, as follows from the equations \eqref{LinEqB}\eqref{LinEqH}, at least in a small neighborhood of the trivial background, there are also other solutions that are not borrowed from YM. (Note that, as follows from the equation \eqref{EoMG}, the appearance of an external Hermitian current $\bm{\Lambda}^a_\text{ext}$ should immediately lead to the appearance of non-zero fields $\bm{N}_a$ and $\bm{G}_{ab}$.)

Finally, if we let the mass of the second gauge field tend to infinity $M\to\infty$, then the equation \eqref{EoMG} leads to $\bm{N}_a = 0$ and, consequently, to $\bm{G}_{ab} = 0$. Then the equation \eqref{EoMF} reduces to $\nabla_b\bm{F}^{ab} = -e^2 \bm{J}^a_\text{ext}$. Thus, \emph{the limit $M\to\infty$ means the transition to the standard YM}.

%%%%%%%%%%%%%%%%%%%%%%%%%%%%%%%%%%%%%%%%%%%%%%%%%%%
\subsection{Redefining the fields}

If we use the Lagrangian in the form \eqref{totalLagrangian}, then there will be coefficients $1/e^2$, $1/\tilde e^2$, and $M^2/4\tilde e^2$ before the kinetic terms $\tr(\hat{\bm D}_{ab}\hat{\bm D}^{ab})/4$, $\tr(\check{\bm D}_{ab}\check{\bm D}^{ab})/4$, and $\tr(\nabla_a\bm{h}\nabla^a\bm{h})/2$
respectively. In order to bring these terms to the standard form with coefficients 1, it is necessary to redefine the perturbations
\begin{equation} \label{Redef1}
\bm{A}_a \mapsto e\bm{A}_a, \quad \bm{B}_a \mapsto \tilde e\bm{B}_a, \quad \bm{h}\mapsto \frac{2\tilde e}{M}\bm{h}.
\end{equation}
The corresponding currents must then change in the reciprocal way:
\begin{equation} \label{Redef2}
\bm{J}_a \mapsto \frac{1}{e}\bm{J}_a, \quad \bm{\Lambda}_a \mapsto \frac{1}{\tilde e}\bm{\Lambda}_a, \quad \bm{E}\mapsto \frac{M}{2\tilde e} \bm{E}.
\end{equation}

Furthermore, it is also convenient to redefine all fields of the $A$-sector to coupling constant $e$, and all fields of the $B$-sector to coupling constant $\tilde e$
\begin{align}
&\bm{\alpha}\mapsto e\bm{\alpha}, \quad \bm{F}_{ab} \mapsto e\bm{F}_{ab}, \label{Redef3} \\
&\bm{\beta}\mapsto \tilde e\bm{\beta}, \quad \bm{G}_{ab} \mapsto \tilde e\bm{G}_{ab}, \quad \bm{N}_a \mapsto \tilde e\bm{N}_a. \label{Redef4}
\end{align} 

Redefinitions \eqref{Redef1}-\eqref{Redef4} must be made in each of the previously obtained formulas. For example, the splits into Hermitian and anti-Hermitian parts will now have the form
\begin{align}
\calA_a &= \tilde e \bm{B}_a - ie \bm{A}_a, & \calF_{ab} &= \tilde e \bm{G}_{ab} - ie \bm{F}_{ab}, \\
\bm{\epsilon} &= \tilde e \bm{\beta} - ie \bm{\alpha}, & \calJ_a &= \frac{1}{\tilde e}\bm{\Lambda}_a - \frac{i}{e}\bm{J}_a.
\end{align}
The fundamental relation \eqref{NtoG}, connecting $\bm{G}_{ab}$ and $\bm{N}_a$, will now read as
\begin{equation}
\bm{G}_{ab} = \nabla_a\bm{N}_b - \nabla_b\bm{N}_a - 2\tilde e [\bm{N}_a, \bm{N}_b]
\end{equation}
and so on.

The mal-YM Lagrangian \eqref{malYMlagrangian}-\eqref{totalLagrangian} after redefining the fields takes form
\begin{equation} \label{FinalActionEq}
\calL_\text{mal-YM} = \frac{1}{4}\tr(\bm{F}_{ab} \bm{F}^{ab} + \bm{G}_{ab}\bm{G}^{ab} + 2M^2 \bm{N}_a \bm{N}^a),
\end{equation}
while equations on the background fields read
\begin{align}
&\nabla_b \bm{F}^{ab} - \tilde e [\bm{N}_b, \bm{F}^{ab}] - ie [\bm{N}_b, \bm{G}^{ab}] = -\bm{J}^a_\text{ext}, \label{EqAfinal} \\
&\nabla_b \bm{G}^{ab} - \tilde e [\bm{N}_b, \bm{G}^{ab}] + i\frac{\tilde e^2}{e} [\bm{N}_b, \bm{F}^{ab}] + M^2\bm{N}^a = \bm{\Lambda}^a_\text{ext}, \label{EqBfinal} \\
&M^2\nabla_a \bm{N}^a + i \frac{e^2+\tilde e^2}{2e} [\bm{G}_{ab}, \bm{F}^{ab}] = \frac{M}{2}\bm{E}_\text{ext}.
\end{align}

%%%%%%%%%%%%%%%%%%%%%%%%%%%%%%%%%%%%%%%%%%%%%%
\section{Gauge fixing} \label{GaugeFixingSec}
%%%%%%%%%%%%%%%%%%%%%%%%%%%%%%%%%%%%%%%%%%%%%%

%%%%%%%%%%%%%%%%%%%%%%%%%%%%%%%%%%%%%%%%%%%%
\subsection{The ``unitary'' $\vect{h}=0$ gauge} \label{EatingGoldstones}

So, in mal-YM, in addition to the usual anti-Hermitian YM potential $\bm{A}_a$, there is also its Hermitian counterpart $\bm{B}_a$, as well as a Goldstone boson (or compensator, or St\"{u}ckelberg field \cite{Ruegg2004}) $\bm{h} = \bm{h}^\dag$. Moreover, since $\bm{h}$ is included in the variation of the action \eqref{FinalActionEq} in combinations \eqref{hDef}
\begin{equation} \label{OomegaNewDef}
\bm{\omega} = \exp(2\tilde e\bm{h}/M), \qquad \bm{\Omega} = \exp(-2\tilde e\bm{h}/M),
\end{equation}
this theory has nonpolynomial interaction, i.e., its Feynman diagrams include vertices of arbitrarily high order in $\bm{h}$.

On the other hand, it has the general $\mathrm{GL}(n, \bbC)$ gauge symmetry \eqref{MatrixTrans}, \eqref{PureGaugePotencalEq}. And, as in other similar cases, this arbitrariness can be exploited to completely absorb the Goldstone boson $\bm{h}$. Indeed, to do this, it is sufficient to perform the non-unitary gauge transformation with the Hermitian parameter
\begin{equation}
\bm{\beta} = -\bm{h}/M,
\end{equation}
which will precisely cancel all the factors $\bm{\omega}$ and $\bm{\Omega}$ \eqref{OomegaNewDef}, i.e., completely eliminate the field $\bm{h}$ and the associated nonpolynomial interactions. This procedure is the fixation of the gauge $\bm{h} = 0$, which is analogous to the standard unitary gauge in theories with spontaneous symmetry breaking via the Higgs mechanism. After this, we still have a residual symmetry in the form of unitary gauge transformations with an anti-Hermitian parameter $\bm{\alpha}$. That is, we are dealing here with spontaneous symmetry breaking $\mathrm{GL}(n, \bbC)\to\mathrm{U}(n)$ (see the Subsection \ref{GaugeSymBreaking} above).

However, this elimination of the $\bm{h}$ field and the associated nonpolynomial interactions come at a very high price. Indeed, if we set $\bm{h} = 0$ in the linearized equation \eqref{LinEqB}, we obtain that $\bm{B}_a$ becomes simply a massive Proca vector field
\begin{equation} \label{ProcaEq}
\big(\delta_a^b(\Box+M^2) + \partial_a\partial^b\big) \bm{B}_b = 0.
\end{equation}
It is well known that theories with a massive Proca field are nonrenormalizable, since the propagator of this field does not decrease in the UV limit $k^2\to\infty$. Indeed, if we rewrite the equation \eqref{ProcaEq} in the momentum representation using the replacement $\partial_a \mapsto ik_a$
\begin{equation}
\big(\delta_a^b(k^2+M^2) + k_a k^b\big) \bm{B}_b = 0,
\end{equation}
then it is easy to see that the propagator in the momentum representation will have the form
\begin{equation}
G_b^c = \frac{1}{k^2+M^2} \Big(\delta_b^c + \frac{k_bk^c}{M^2}\Big) \xrightarrow[k^2\to\infty]{} \frac{k_bk^c}{k^2M^2}.
\end{equation}

On the other hand, it is equally well known that this problem can be circumvented and gauge fields can be made massive, due to spontaneous symmetry breaking via the Higgs mechanism. Since we have a largely analogous situation with spontaneous symmetry breaking in mal-YM, one might hope that the same trick will work in this case as well. And, as we will see in the next subsection, this turns out to be indeed the case.

%%%%%%%%%%%%%%%%%%%%%%%%%%%%%%%%%%%%%%%%%%%%
\subsection{The Feynman-'t Hooft gauge} \label{xiGauge}

As we saw above \eqref{TrivLinAction}-\eqref{LinEqH}, on the trivial background $\bm{F}_{ab} = 0$, $\bm{N}_a = 0$, the quadratic part of the action and the resulting linearized EoMs for the field $\bm{A}_a$ and the fields $\bm{B}_a$, $\bm{h}$ are separated, so they can be considered independently.

The field $\bm{A}_a$ is the usual YM potential. The standard $R_\xi$ gauge for it is introduced by adding to the quadratic action
\begin{equation}
\calL_A^{(2)} = \frac{1}{4}\tr(\partial_a\bm{A}_b - \partial_b\bm{A}_a)(\partial^a\bm{A}^b - \partial^b\bm{A}^a)
\end{equation}
the gauge-fixing term
\begin{equation}
\calL_A^\text{gf} = \frac{1}{2\xi} \tr(\partial_a\bm{A}^a)^2.
\end{equation}
This leads to the linearized equation
\begin{equation}
\Big(\delta_a^b \Box + \tfrac{\xi - 1}{\xi}\partial_a\partial^b\Big) \bm{A}_b = 0,
\end{equation}
which corresponds to the following propagator in the momentum representation
\begin{equation}
G_b^c = \frac{1}{k^2}\Big(\delta_b^c + (\xi-1)\frac{k_bk^c}{k^2}\Big).
\end{equation}
This expression becomes the most simple in the special case of the Feynman gauge $\xi=1$.

In contrast, the fields $\bm{B}_a$ and $\bm{h}$ form a St\"uckelberg type system with a quadratic Lagrangian [after redefinition \eqref{Redef1}]
\begin{align} \label{BhAction}
\calL_{Bh}^{(2)} &= \frac{1}{4}\tr(\partial_a\bm{B}_b - \partial_b\bm{B}_a)(\partial^a\bm{B}^b - \partial^b\bm{B}^a) \nonumber \\
&+\frac{1}{2}\tr(\partial_a\bm{h} - M\bm{B}_a)(\partial^a\bm{h} - M\bm{B}^a),
\end{align}
which is invariant under non-unitary gauge transformations
\begin{equation}
\bm{B}_a \mapsto \bm{B}_a + \partial_a\bm{\beta}, \qquad \bm{h} \mapsto \bm{h} + M\bm{\beta}.
\end{equation}

For it, we can introduce a two-parameter gauge, which we will call the generalized Feynman-'t Hooft gauge, using the following term
\begin{equation} \label{FeynmanTHooft}
\calL_{Bh}^\text{gf} = \frac{1}{2\xi} \tr\Big(\partial_a\bm{B}^a - m\bm{h}\Big)^2.
\end{equation}
This leads to the system of coupled equations
\begin{align}
&\big(\delta_a^b (\Box + M^2) + \tfrac{\xi - 1}{\xi} \partial_a\partial^b\big) \bm{B}_b +\big(\tfrac{m}{\xi} - M\big) \partial_a\bm{h} = 0, \\
&\big(\Box + \tfrac{m^2}{\xi}\big) \bm{h} + \big(M - \tfrac{m}{\xi}\big) \partial_a \bm{B}^a = 0.
\end{align}

Remarkably, by choosing $m = \xi M$, this system decouples---the fields $\bm{B}_a$ and $\bm{h}$ become independent of each other. This happens because the cross terms in the expressions \eqref{BhAction} and \eqref{FeynmanTHooft} add up to the total derivative. If we additionally set $\xi=1$, the equations take a particularly simple form:
\begin{equation}
(\Box + M^2)\bm{B}_a = 0, \qquad (\Box + M^2)\bm{h} = 0,
\end{equation}
i.e., they describe two independent fields---a scalar and a vector---of the same mass $M^2$. We will call this specific choice of parameters $\xi=1$, $m=M$ simply the Feynman-'t Hooft gauge. It is very convenient to use in calculations, and to use the generalized gauge with arbitrary parameters $\xi$, $m$ for the results verification.

Thus, using the Feynman-'t Hooft gauge, one can circumvent the problem of the non-decreasing propagator of the massive Proca field, which arises in the ``unitary'' gauge $\bm{h} = 0$. However, the price for this is that the St\"uckelberg field $\bm{h}$ must be treated as a dynamical field carrying degrees of freedom and participating in nonpolynomial interactions.

%%%%%%%%%%%%%%%%%%%%%%%%%%%%%%%%%%%%%%%%%%%%%%
\section{Interactions} \label{InteractionsSec}
%%%%%%%%%%%%%%%%%%%%%%%%%%%%%%%%%%%%%%%%%%%%%%

In this section, we consider interactions in mal-YM. Although we can explicitly write out the interaction vertices for an arbitrary given background $\bm{F}_{ab}$ and $\bm{N}_{a}$ (and such expressions are needed to calculate the quantum effective action in the background field formalism), they are rather cumbersome and uninformative. Therefore, we simply present them for reference in the Appendix~\ref{VertApp}. Here, as above in the Subsection~\ref{LinearizedEoMsSubsec}, we restrict ourselves to the case of the trivial background $\bm{F}_{ab}=0$ and $\bm{N}_{a}=0$.

We will proceed as follows: first, in Subsection~\ref{ParitySubsec}, we note that, on a trivial background, the system has an additional symmetry, which turns out to be associated with a conserved quantity. Then, in Subsection~\ref{VerticesWithH}, we expand the Lagrangian \eqref{FinalActionEq} in powers of the fields $\bm{A}_a$ and $\bm{B}_a$, keeping the field $\bm{h}$ inside the Hermitian matrices $\bm{\omega}$ and $\bm{\Omega}$ \eqref{OomegaNewDef}. The corresponding term in the Lagrangian that has power $p$ in the fields $\bm{A}_a$ and power $q$ in the fields $\bm{B}_a$ we will denote by $\calL_{pq}$. In Subsection~\ref{VerticesStructure}, we also expand in powers of the fields $\bm{h}$ and analyze the first few terms of this expansion. The corresponding term in the action, which also has degree $r$ over the fields $\bm{h}$, we will denote as $\calL_{pqr}$.

%%%%%%%%%%%%%%%%%%%%%%%%%%%%%%%%%%%%%%%%%%%%%%%%%%%
\subsection{St\"{u}ckelberg parity} \label{ParitySubsec}

Now we start with a trivial background, i.e., with a Hermitian form $g_{\alpha\beta'}$ and derivative $\partial_a$ such that $\bm{F}_{ab}[\partial, g] = 0$ and $\bm{N}_a[\partial, g] = 0$. Then, using perturbations $\bm{A}_a$, $\bm{B}_a$, and $\bm{h}$, we pass to some new Hermitian form $\tilde g_{\alpha\beta'}$ and derivative $\tilde\nabla_a$. From the transformations \eqref{Ntransformation} and \eqref{FinalFGtrans3}-\eqref{FinalFGtrans4} we obtain the following expressions for them:
\begin{align}
\bm{N}_a &= \frac{1}{2} \big(\calA_a + \bm{\Omega}\calA_a^\dag \bm{\omega} - \bm{\Omega}\partial_a\bm{\omega} \big), \label{TrTr1} \\
\bm{G}_{ab} &= \frac{1}{2} \big(\bm{G}'_{ab} +\bm{\Omega}\bm{G}'_{ab}\bm{\omega} \big) - \frac{i}{2} \big(\bm{F}'_{ab} -\bm{\Omega}\bm{F}'_{ab}\bm{\omega} \big), \label{TrTr2} \\
\bm{F}_{ab} &= \frac{1}{2} \big(\bm{F}'_{ab} +\bm{\Omega}\bm{F}'_{ab}\bm{\omega} \big) + \frac{i}{2} \big(\bm{G}'_{ab} -\bm{\Omega}\bm{G}'_{ab}\bm{\omega} \big). \label{TrTr3}
\end{align}

Consider a transformation that changes the sign of the St\"{u}ckelberg sector fields
\begin{equation} \label{StuckelbergParity}
\bm{A}_a\mapsto\bm{A}_a, \qquad \bm{B}_a\mapsto -\bm{B}_a, \qquad \bm{h}\mapsto -\bm{h}.
\end{equation}

From \eqref{checkDC}-\eqref{bmCdef} and \eqref{OomegaNewDef} it is immediately clear that the quantities defined therein transform as follows:
\begin{align}
\check{\bm D}_{ab} &\mapsto -\check{\bm D}_{ab}, & \bm{C}_{ab} &\mapsto -\bm{C}_{ab}, & \bm{\omega} &\leftrightarrow \bm{\Omega}, \label{DCtrans1} \\
\hat{\bm D}_{ab} &\mapsto \hat{\bm D}_{ab}, & \hat{\bm C}_{ab} &\mapsto \hat{\bm C}_{ab}, & \check{\bm C}_{ab} &\mapsto \check{\bm C}_{ab}. \label{DCtrans2}
\end{align}
Then from \eqref{FinalFGtrans1}-\eqref{FinalFGtrans2} it follows that
\begin{equation}
\bm{F}'_{ab} \mapsto \bm{F}'_{ab}, \qquad \bm{G}'_{ab} \mapsto -\bm{G}'_{ab},
\end{equation}
and, finally, from \eqref{TrTr1}-\eqref{TrTr3} we obtain the following background field transformations
\begin{equation} \label{BackgroundSymmetry}
\bm{F}_{ab} \mapsto \bm{\omega}\bm{F}_{ab}\bm{\Omega}, \quad \bm{G}_{ab} \mapsto -\bm{\omega}\bm{G}_{ab}\bm{\Omega}, \quad \bm{N}_a \mapsto -\bm{\omega}\bm{N}_a\bm{\Omega}.
\end{equation}

However, it is easy to see that the background field transformations \eqref{BackgroundSymmetry} (for any mutually inverse matrices $\bm{\omega}$ and $\bm{\Omega}$) are a symmetry of the Lagrangian $\calL_\text{mal-YM}$ \eqref{FinalActionEq} under consideration. Consequently, variations of the action over the trivial background will also be invariant under the transformations of perturbations \eqref{StuckelbergParity}. This symmetry will correspond to a certain conserved quantity, which we will call the \emph{St\"{u}ckelberg parity}.

In the next subsection, we write out the terms $\calL_{pq}$ obtained by expanding the Lagrangian in powers of the fields $\bm{A}_a$ and $\bm{B}_a$ (but not in the field $\bm{h}$). Each of these terms individually will also be invariant under the transformations \eqref{DCtrans1}-\eqref{DCtrans2}. When we further expand in powers of $\bm{h}$, we obtain terms $\calL_{pqr}$, which correspond to the vertices of the theory with $p$ outgoing legs corresponding to $\bm{A}_a$, $q$ outgoing $\bm{B}_a$-legs (there can be no more than four of them in total, $p+q\le4$), and $r$ outgoing $\bm{h}$-legs (there can be an unlimited number of them---this theory is nonpolynomial). However, each of these terms must be symmetric under the transformation \eqref{StuckelbergParity}. From this, in turn, it will follow that \emph{all vertices with an odd number of outgoing $B$-sector fields' legs are equal to zero}
\begin{equation}
\calL_{pqr} = 0 \quad\text{for}\quad q+r = 2k+1.
\end{equation}

This means that in pure mal-YM, particles of an arbitrary mass $M$, described by the $B$-sector fields $\bm{B}_a$ and $\bm{h}$, can be produced in pairs through collisions of YM bosons and annihilate with each other, but cannot decay on their own. Thus, the St\"{u}ckelberg parity is simply $(-1)^{n_B}$, where $n_B$ is the number of St\"{u}ckelberg particles in the system.

Note that if this was always the case, then $B$-sector particles could be considered potential candidates for WIMPs. However, as we will see below in Subsection~\ref{ScalarSubsec}, any interaction with matter leads to violation of St\"{u}ckelberg parity conservation. In this case, the corresponding particles will decay into particles of ordinary matter of smaller masses. This is not surprising, since even in pure mal-YM, conservation of St\"{u}ckelberg parity is a consequence of the arbitrary choice of the Lagrangian $\calL_\text{mal-YM}$~\eqref{FinalActionEq}. If we were to add, say, the term $\tr(\bm{G}_{ab}\bm{F}^{ab})$, obviously, it would not be conserved by the transformations~\eqref{BackgroundSymmetry}.

%%%%%%%%%%%%%%%%%%%%%%%%%%%%%%%%%%%%%%%%%%%%%%%%%%%
\subsection{Expansion in powers of $\vect{A}_a$ and $\vect{B}_a$} \label{VerticesWithH}

Substituting expressions \eqref{TrTr1}-\eqref{TrTr3} into the Lagrangian \eqref{FinalActionEq} and expanding the result in powers of $\vect{A}_a$ and $\vect{B}_a$, we obtain 15 terms $\calL_{pq}$, where $p+q\le4$. The terms of orders 0 and 1 look extremely simple:
\begin{align}
\calL_{00} &= \frac{M^2}{8\tilde e^2} \tr\big(\bm{\Omega}\partial_a\bm{\omega}\bm{\Omega}\partial^a\bm{\omega}\big), \label{CalL00h} \\
\calL_{10} &= \frac{ieM^2}{4\tilde e^2} \tr\big( [\bm{A}^a, \bm{\Omega}]\partial_a\bm{\omega}\big), \\
\calL_{01} &= -\frac{M^2}{4\tilde e} \tr\big( \{\bm{B}^a, \bm{\Omega}\}\partial_a\bm{\omega}\big),
\end{align}
where square brackets denote the commutator, and curly brackets denote the anticommutator. Also, from here on, we assume that if there are no parentheses, then the derivative $\partial_a$ acts only on the term immediately following it.

We also introduce the following abbreviated notations:
\begin{equation} \label{Epm}
e_+^2 = \tilde e^2 + e^2, \qquad e_-^2 = \tilde e^2 - e^2.
\end{equation}
Then the terms of the order two have the form:
\begin{align}
\calL_{20} &= \frac{e^2M^2}{4\tilde e^2} \tr\big(\bm{A}_a\bm{\Omega}\bm{A}^a\bm{\omega} - \bm{A}_a\bm{A}^a\big) \nonumber \\
&+ \frac{1}{8\tilde e^2} \tr\big(e_+^2 \hat{\bm D}_{ab}\bm{\Omega}\hat{\bm D}^{ab}\bm{\omega} + e_-^2 \hat{\bm D}_{ab}\hat{\bm D}^{ab}\big), \\
\calL_{02} &= \frac{M^2}{4} \tr\big(\bm{B}_a\bm{\Omega}\bm{B}^a\bm{\omega} + \bm{B}_a\bm{B}^a\big) \nonumber \\
&+ \frac{1}{8e^2} \tr\big(e_+^2 \check{\bm D}_{ab}\bm{\Omega}\check{\bm D}^{ab}\bm{\omega} - e_-^2 \check{\bm D}_{ab}\check{\bm D}^{ab}\big), \\
\calL_{11} &= \frac{ieM^2}{4\tilde e} \tr\big(\bm{B}_a\bm{\Omega}\bm{A}^a\bm{\omega} - \bm{A}_a\bm{\Omega}\bm{B}^a\bm{\omega}\big) \nonumber \\
&+ \frac{ie_+^2}{8e\tilde e} \tr\big(\check{\bm D}_{ab}\bm{\Omega}\hat{\bm D}^{ab}\bm{\omega} - \hat{\bm D}_{ab}\bm{\Omega}\check{\bm D}^{ab}\bm{\omega} \big).
\end{align}

For terms of the order three, we obtain:
\begin{align}
\calL_{30} &= \frac{e}{8\tilde e^2} \tr\Big(e_+^2\big(\hat{\bm C}_{ab}\bm{\Omega}\hat{\bm D}^{ab}\bm{\omega} + \hat{\bm D}_{ab}\bm{\Omega}\hat{\bm C}^{ab}\bm{\omega}\big) \nonumber \\
&- 2e_-^2 \hat{\bm C}_{ab}\hat{\bm D}^{ab}\Big), \\
\calL_{03} &= \frac{i\tilde ee_+^2}{8e^2} \tr\big(\check{\bm D}_{ab}\bm{\Omega}\check{\bm C}^{ab}\bm{\omega} - \check{\bm C}_{ab}\bm{\Omega}\check{\bm D}^{ab}\bm{\omega}\big), \\
\calL_{21} &= \frac{ie_+^2}{8\tilde e} \tr\big(\hat{\bm C}_{ab}\bm{\Omega}\check{\bm D}^{ab}\bm{\omega} - \check{\bm D}_{ab}\bm{\Omega}\hat{\bm C}^{ab}\bm{\omega} \nonumber \\
&+ \hat{\bm D}_{ab}\bm{\Omega}\bm{C}^{ab}\bm{\omega} - \bm{C}_{ab}\bm{\Omega}\hat{\bm D}^{ab}\bm{\omega}\big), \\
\calL_{12} &= \frac{1}{8 e} \tr\Big(e_+^2 \big(\check{\bm C}_{ab}\bm{\Omega}\hat{\bm D}^{ab}\bm{\omega} + \hat{\bm D}_{ab}\bm{\Omega}\check{\bm C}^{ab}\bm{\omega} \nonumber \\
&- \check{\bm D}_{ab}\bm{\Omega}\bm{C}^{ab}\bm{\omega} - \bm{C}_{ab}\bm{\Omega}\check{\bm D}^{ab}\bm{\omega}\big) \nonumber \\
&+ 2e_-^2\big(\check{\bm D}_{ab} \bm{C}^{ab} + \check{\bm C}_{ab}\hat{\bm D}^{ab}\big)\Big).
\end{align}

Finally, the five order four terms are:
\begin{align}
\calL_{40} &= \frac{e^2}{8\tilde e^2} \tr\big(e_+^2 \hat{\bm C}_{ab}\bm{\Omega}\hat{\bm C}^{ab}\bm{\omega} + e_-^2 \hat{\bm C}_{ab}\hat{\bm C}^{ab}\big), \\
\calL_{04} &= \frac{\tilde e^2}{8e^2} \tr\big(e_+^2 \check{\bm C}_{ab}\bm{\Omega}\check{\bm C}^{ab}\bm{\omega} + e_-^2 \check{\bm C}_{ab}\check{\bm C}^{ab}\big), \\
\calL_{31} &= \frac{iee_+^2}{8\tilde e} \tr\big(\bm{C}_{ab}\bm{\Omega}\hat{\bm C}^{ab}\bm{\omega} - \hat{\bm C}_{ab}\bm{\Omega}\bm{C}^{ab}\bm{\omega}\big), \\
\calL_{13} &= \frac{i\tilde ee_+^2}{8e} \tr\big(\check{\bm C}_{ab}\bm{\Omega}\bm{C}^{ab}\bm{\omega} - \bm{C}_{ab}\bm{\Omega}\check{\bm C}^{ab}\bm{\omega}\big), \\
\calL_{22} &= \frac{1}{8} \tr\Big(e_+^2\big(\bm{C}_{ab}\bm{\Omega}\bm{C}^{ab}\bm{\omega} - \check{\bm C}_{ab}\bm{\Omega}\hat{\bm C}^{ab}\bm{\omega} \label{CalL22h} \nonumber \\
&- \hat{\bm C}_{ab}\bm{\Omega}\check{\bm C}^{ab}\bm{\omega}\big) - e_-^2\big(\bm{C}_{ab}\bm{C}^{ab} + 2\hat{\bm C}_{ab}\check{\bm C}^{ab}\big)\Big).
\end{align}

It can be easily verified that each of the 15 terms given above indeed remains unchanged under transformations~\eqref{DCtrans1}-\eqref{DCtrans2}.

%%%%%%%%%%%%%%%%%%%%%%%%%%%%%%%%%%%%%%%%%%%%%%%%%%%
\subsection{Interaction vertices on the trivial background} \label{VerticesStructure}

Now, expanding the terms \eqref{CalL00h}-\eqref{CalL22h} into an infinite series in powers of the field $\bm{h}$ using the definition~\eqref{OomegaNewDef}, we obtain expressions corresponding to each of the vertices of our theory. Since mal-YM has an infinite number of vertices with arbitrarily high powers of the field $\bm{h}$, we will write out only the first few of them.

Mal-YM contains the standard YM vertices of order~3 and~4:
\begin{equation}\label{OrdThreeVericesOrdinaryYM}
\calL_{300} = -\frac{e}{2}\tr(\hat{\bm C}_{ab}\hat{\bm D}^{ab}), \qquad \calL_{400} = \frac{e^2}{4}\tr(\hat{\bm C}_{ab}\hat{\bm C}^{ab}).
\end{equation}
However, apart from~\eqref{OrdThreeVericesOrdinaryYM}, one additional third-order vertex appears, with one outgoing leg corresponding to the field $\bm{A}_a$ and two outgoing legs of field~$\bm{B}_a$:
\begin{equation}
\calL_{120} = -\frac{e}{2}\tr(\bm{C}_{ab}\check{\bm D}^{ab}) + \frac{\tilde e^2}{2e} \tr(\check{\bm C}_{ab}\hat{\bm D}^{ab}).
\end{equation}
There are also two new fourth-order vertices: one with two outgoing legs of the fields $\bm{A}_a$ and $\bm{B}_a$, and one with four legs of the field $\bm{B}_a$:
\begin{align}
\calL_{220} &= \frac{e^2}{4}\tr(\bm{C}_{ab} \bm{C}^{ab}) - \frac{\tilde e^2}{2} \tr(\check{\bm C}_{ab}\hat{\bm C}^{ab}), \\
\calL_{040} &= \frac{\tilde e^4}{4e^2}\tr(\check{\bm C}_{ab}\check{\bm C}^{ab}).
\end{align}

This is as it should be due to the above-mentioned conservation of St\"{u}ckelberg parity in pure mal-YM: all the vertices listed above are even in the field $\bm{B}_a$, while the odd vertices of order 3 and 4 $\calL_{210}$, $\calL_{030}$, $\calL_{310}$, and $\calL_{130}$ turn out to be equal to zero. Conversely, the terms linear in $\bm{h}$, are all odd in $\bm{B}_a$:
\begin{align}
\calL_{211} &= \frac{ie_+^2}{2M} \tr\big(([\hat{\bm C}_{ab}, \check{\bm D}^{ab}] - [\bm{C}_{ab}, \hat{\bm D}^{ab}]) \bm{h} \big), \\
\calL_{031} &= \frac{i\tilde e^2e_+^2}{2e^2M} \tr\big([\check{\bm D}_{ab}, \check{\bm C}^{ab}] \bm{h}\big), \\
\calL_{311} &= \frac{iee_+^2}{2M} \tr\big(\bm{C}_{ab}, \hat{\bm C}^{ab}] \bm{h}\big), \\
\calL_{131} &= \frac{i\tilde e^2e_+^2}{2eM} \tr\big([\check{\bm C}_{ab}, \bm{C}^{ab}] \bm{h}\big),
\end{align}
while the vertices even in $\bm{B}_a$ are all equal to zero.

Of course, we can write out other vertices in exactly the same way. Here are four more simple examples:
\begin{align}
\calL_{102} &= ie \tr\big([\bm{h}, \bm{A}^a] \partial_a\bm{h}\big), \\
\calL_{004} &= \frac{\tilde e^2}{3M^2} \tr\big((\bm{h}^2\partial_a\bm{h} - \bm{h}\partial_a\bm{h}\bm{h})\partial^a\bm{h}\big), \\
\calL_{013} &= \frac{2\tilde e^2}{3M} \tr\big((2\bm{h}\bm{B}^a\bm{h} - \bm{h}^2\bm{B}^a - \bm{B}^a\bm{h}^2)\partial_a\bm{h}\big), \\
\calL_{111} &= ieM \tr\big([\bm{B}_a, \bm{A}^a]\bm{h}\big) + \frac{ie_+^2}{2eM} \tr\big([\check{\bm D}_{ab}, \hat{\bm D}^{ab}]\bm{h}\big).
\end{align}

%%%%%%%%%%%%%%%%%%%%%%%%%%%%%%%%%%%%%%%%%%%%%%
\subsection{Interaction with a scalar field} \label{ScalarSubsec}

Finally, let us briefly consider the interaction of mal-YM with gauge-charged (i.e., carrying an internal color index) matter fields. We will do this using the simplest example of a scalar field, for which it is convenient to introduce the following notation:
\begin{align}
&\bm{\varphi} \cong \varphi^\alpha, \qquad \bm{\varphi}^\dag \cong \bar\varphi_\alpha = g_{\alpha\beta'}\bar\varphi^{\beta'}, \\
&\bm{\varphi}^\dag \bm{\varphi} = \bar\varphi_\alpha\varphi^\alpha = g_{\alpha\beta'}\varphi^\alpha \bar\varphi^{\beta'}.
\end{align}
However, when using index-free notation here, as elsewhere, one must not forget that after relaxation of the covariant constancy condition~\eqref{CovConstYM}, the operation of Hermitian conjugation no longer commutes with covariant differentiation. It is easy to obtain the following analogue of the formula~\eqref{DerHermEq}:
\begin{equation}
(\nabla_a\bm{\varphi})^\dag = \nabla_a(\bm{\varphi}^\dag) + 2\tilde e \bm{\varphi}^\dag\bm{N}_a.
\end{equation}

Then the Lagrangian describing a scalar gauge-charged field minimally interacting with mal-YM can be chosen as
\begin{equation} \label{PhiLag}
\calL_\varphi = \frac{1}{2} (\nabla_a\bm{\varphi})^\dag \nabla^a \bm{\varphi} + P\left(\bm{\varphi}^\dag \bm{\varphi}\right),
\end{equation}
where $P$ is a certain self-interaction potential of the field $\bm{\varphi}$.

If we make a transformation to a new connection $\tilde\nabla_a$ using the total potential $\calA_a$ and~\eqref{RaznDiff}, and to a new Hermitian form $\tilde g_{\alpha\beta'}$ using the matrix $\bm{\omega}$ and~\eqref{gTrans}, then this Lagrangian becomes
\begin{equation} \label{tildePhiLag}
\tilde\calL_\varphi = \frac{1}{2} \big((\nabla_a+\calA_a)\bm{\varphi}\big)^\dag \bm{\omega} (\nabla^a + \calA_a)\bm{\varphi} + P\left(\bm{\varphi}^\dag \bm{\omega} \bm{\varphi}\right).
\end{equation}

Then the gauge currents generated by the Lagrangian \eqref{PhiLag} and defined by the formulas \eqref{totalCurrent}-\eqref{hSource} will have the form
\begin{align}
&\calJ_a = -\bm{\varphi}\otimes (\nabla_a\bm{\varphi})^\dag, \qquad \calJ_a^\dag = -\nabla_a\bm{\varphi}\otimes \bm{\varphi}^\dag, \\
&\bm{\Lambda}_a = -\frac{\tilde e}{2} \big(\nabla_a\bm{\varphi}\otimes\bm{\varphi}^\dag + \bm{\varphi}\otimes (\nabla_a\bm{\varphi})^\dag \big), \label{CurrentPhiB} \\
&\bm{J}_a = \frac{ie}{2} \big(\nabla_a\bm{\varphi}\otimes \bm{\varphi}^\dag - \bm{\varphi}\otimes (\nabla_a\bm{\varphi})^\dag\big), \label{CurrentPhiA} \\
&\bm{E} = \frac{\tilde e}{M} \big(\nabla_a\bm{\varphi} \otimes (\nabla^a\bm{\varphi})^\dag + 2P' \bm{\varphi}\otimes \bm{\varphi}^\dag\big),
\end{align}
where it is implied that ``the product of a column and a row yields a matrix,'' for example, $\bm{\varphi}\otimes \bm{\varphi}^\dag \cong \varphi^\alpha\bar\varphi_\beta$.

It is convenient to write the source to the field $\bm{\varphi}$ in the form
\begin{equation}
\bm{\rho} \cong \rho^\alpha = g^{\alpha\beta'} \frac{\delta S}{\delta \bar\varphi^{\beta'}} \cong \big(\frac{\mathbf{1}}{2}\Box + \tilde e \bm{N}^a\nabla_a  + \mathbf{1}P'\big) \bm{\varphi}.
\end{equation}
Then the complete system of classical equations for background fields describing mal-YM together with the charged scalar $\bm{\varphi}$ will have the following form: these are the equations \eqref{EqAfinal}-\eqref{EqBfinal}, into which we substitute \eqref{CurrentPhiB}-\eqref{CurrentPhiA} instead of external sources $\bm{\Lambda}^a_\mathrm{ext}$ and $\bm{J}^a_\mathrm{ext}$, plus the equation $\bm{\rho} = 0$.

From the expression \eqref{tildePhiLag} it is not difficult to obtain all interaction vertices. If we do not consider vertices involving the field $\bm{h}$, then the situation is quite simple: together with the usual third-order vertex $\tr(\bm{J}_a\bm{A}^a)$, in which the propagator of the field $\bm{A}^a$ is attached to the line of the scalar field $\bm{\varphi}$, the vertex $-\tr(\bm{\Lambda}_a\bm{B}^a)$ will appear. And to the fourth-order vertex $\bm{\varphi}^\dag\bm{A}_a\bm{A^a}\bm{\varphi}$, two more vertices involving the field $\bm{B}^a$ are added, namely $i\bm{\varphi}^\dag[\bm{A}_a, \bm{B^a}]\bm{\varphi}$ and $\bm{\varphi}^\dag\bm{B}_a\bm{B^a}\bm{\varphi}$. The situation is complicated by the fact that we also have an infinite number of vertices involving the field $\bm{h}$, arising from the expansion of the matrix $\bm{\omega}$ in \eqref{tildePhiLag}. In other words, the field $\bm{h}$ can attach itself in any power to any line of the field $\bm{\varphi}$ and to any vertex that includes $\bm{\varphi}$.

Note that this means that the interaction with a scalar, described by the Lagrangian \eqref{PhiLag}, violates the St\"uckelberg parity we introduced earlier in the Subsection~\ref{ParitySubsec}. A similar situation should also occur for interactions with fermions. This means that despite the presence of parity in pure mal-YM, particles in the St\"uckelberg  sector will decay into particles of ordinary matter of lower mass, and thus are an unlikely dark matter candidate.

%%%%%%%%%%%%%%%%%%%%%%%%%%%%%%%%%%%%%%%%%%%%%%%%
\section{Conclusion} \label{ConclusionSec}

To summarize: previously in \cite{Wachowski24a}, we first proposed a natural generalization of $\mathrm{U}(n)$ Yang-Mills (YM) theory in which the condition of covariant constancy of the Hermitian form \eqref{CovConstYM} is no longer satisfied, and thus the connection $\nabla_a$ and $g_{\alpha\beta'}$ become independent field variables. Since it is based on an analogy with metric-affine gravity (MAG), we chose to call this generalization of Yang-Mills theory ``metric-affine-like''---mal-YM. Although we initially viewed mal-YM as a kind of toy model that might allow us to better understand the properties of MAG, it turns out that it has very interesting properties in its own right.

In pure mal-YM there are two sectors: the Yang-Mills $A$-sector consists of the usual potential $\bm{A}_a$ and field strength tensor $\bm{F}_{ab}$. However, there is also an additional $B$-sector, which interacts non-trivially with standard YM field. This sector consists of the fields $\bm{h}$ and $\bm{B}_a$, $\bm{N}_a$ and $\bm{G}_{ab}$ (which are also all Hermitian matrices). In essence, the $B$-sector is a non-Abelian generalization of the well-known St\"{u}ckelberg theory \cite{Ruegg2004}---to our knowledge, such a generalization has also not been obtained previously. The fields of the St\"{u}ckelberg sector can be given an arbitrary mass $M$; as it tends to infinity $M\to\infty$, the associated degrees of freedom will effectively freeze out, which recovers the standard YM.

In this paper, we have analyzed mal-YM in detail at the classical level. A much more interesting and still open question is whether this theory is well-behaved at the quantum level, and, in particular, whether it is renormalizable or not. Generally speaking, problems with the renormalizability of a theory arise from two sources: either from insufficiently rapid decay of the propagators or from the presence of vertices with negative coupling dimensions. At a first glance, mal-YM seems to have problems of both kinds: it contains a massive vector boson (and the Proca field propagator is non-decreasing) and nonpolynomial interactions with vertices of arbitrarily high order and, accordingly, arbitrarily large negative coupling dimensions. Therefore, it might even seem that mal-YM is simply pathological and meaningless.

However, the situation turns out to be much more subtle. In the theory of electroweak interactions, vector bosons obtain masses through spontaneous symmetry breaking via the Higgs mechanism. And in mal-YM the situation is almost the same: in this theory, the Hermitian form $g_{\alpha\beta'}$ is also a ``Higgs field,'' spontaneously breaking the symmetry from $\mathrm{GL}(n, \bbC)$ to $\mathrm{U}(n)$. Accordingly, this broken $\mathrm{GL}(n, \bbC)$ symmetry can be used to fight the above mentioned problems. As we showed in Sec.~\ref{GaugeFixingSec}, in the $\bm{h}=0$ gauge, similar to the unitary gauge, all nonpolynomial interactions associated with the field $\bm{h}$ vanish (however, then $\bm{B}_a$ becomes a Proca field with an ill-behaved propagator). On the other hand, in the Feynman-'t~Hooft gauge, all propagators behave well, but vertices of arbitrarily high order appear. Our hope is that, due to the broken $\mathrm{GL}(n, \bbC)$ gauge symmetry, the divergences associated with these vertices will cancel each other out. However, this is a very subtle and delicate issue that requires further careful study.

Even if mal-YM turns out to be a pathological theory at the quantum level, this would already be a very interesting result, since, firstly, similar (and even more serious) problems would also arise in MAG. And, secondly, it would improve our understanding the standard YM, which satisfies the covariant constancy condition~\eqref{CovConstYM}---why is it structured like this and not some other way?

However, if mal-YM turns out to be a physically admissible theory, this would allow us to consider it as a potential candidate for an extension of the Standard Model. The important point is that we are not able to distinguish mal-YM from YM if the currently accessible energy scale is significantly smaller than the mass $M$. Mal-YM is a broader theory with a larger number of parameters, and therefore it can potentially describe a wider range of phenomena. In any case, it is already possible to attempt to obtain a lower limit on the mass $M$ from data of collider experiments and observational cosmology. One might wonder whether mal-YM might describe some deviations from the Standard Model and consider which phenomena would provide biggest differences between the two theories.

%%%%%%%%%%%%%%%%%%%%%%%%%%%%%%%%%
\section*{Acknowledgments}

The author expresses deep gratitude to A.\,E.~Kalugin, A.\,O.~Barvinsky and D.\,V.~Nesterov for stimulating discussions, and I.\,M.~Vysotsky for discussing the problem of non-decreasing propagator in Proca theory.

%%%%%%%%%%%%%%%%%%%%%%%%%%%%%%%%%%%%%%%%%%%%%%%%
\appendix
\section{On general linear connection} \label{CurvApp}
%%%%%%%%%%%%%%%%%%%%%%%%%%%%%%%%%%%%%%%%%%%%%%%%

\subsection{Derivation of curvature transformations and Bianchi identities}

In this appendix, for reference, we provide the derivations of the transformation laws for torsion \eqref{TorsionPreobr} and curvatures \eqref{FPreobr}-\eqref{tildeFPreobr}, and the Bianchi identities \eqref{AlgBianSym}-\eqref{DifBian}. We also give the corresponding generalizations for the case of nonzero torsion.

Let us start with the transformation law for torsion \eqref{TorsionPreobr}. To derive it, we apply the commutator of covariant derivatives $2\nabla_{[a}\nabla_{b]}$ to an arbitrary scalar $f$:
\begin{align}
2\tilde\nabla_{[a}\tilde\nabla_{b]}f &= T_{ab}{}^c[\tilde\nabla]\tilde\nabla_c f = T_{ab}{}^c[\tilde\nabla]\nabla_c f \nonumber \\
= 2\tilde\nabla_{[a}\nabla_{b]}f &= 2\left(\nabla_{[a}\nabla_{b]} - \scrA_{[ab]}{}^c \nabla_c\right) f \nonumber \\
&= \left(T_{ab}{}^c[\nabla] - 2\scrA_{[ab]}{}^c\right) \nabla_c f.
\end{align}
Here we used the general connection transformation \eqref{RaznDiff} (taking into account that derivatives coincide on scalars $\tilde\nabla_a f = \nabla_a f$) and twice the definition of torsion \eqref{DefT}. Since $\nabla_a f$ is locally an arbitrary covector, this immediately implies the relation \eqref{TorsionPreobr}.

To derive the transformation law for curvature in the bundle $\calF_{ab}$ we again act with the commutator of covariant derivatives $\nabla_{[a}\nabla_{b]}$, but now not upon a scalar, but rather upon a quantity $\bm{\psi} \cong \psi^\alpha$. Applying the rule \eqref{RaznDiff} twice, we get:
\begin{align}
&\tilde\nabla_{[a}\tilde\nabla_{b]} \bm{\psi} = \left(\nabla_{[a}\tilde\nabla_{b]} + \calA_{[a}\tilde\nabla_{b]} - \scrA_{[ab]}{}^c\tilde\nabla_c\right)\bm{\psi} \nonumber \\
&= \big(\nabla_{[a}\nabla_{b]} + (\nabla_{[a}\calA_{b]}) + \calA_{[a}\calA_{b]} - \scrA_{[ab]}{}^c\tilde\nabla_c \big) \bm{\psi}.
\end{align}
On the other hand, by definition \eqref{DefF} we have
\begin{equation}
2\nabla_{[a}\nabla_{b]} \bm{\psi} = \left(\calF_{ab}[\nabla] + T_{ab}{}^c[\nabla]\nabla_c\right) \bm{\psi}
\end{equation}
and a similar relation for the connection $\tilde\nabla_a$. Combining these formulas and using \eqref{TorsionPreobr}, we get
\begin{equation} \label{genFtrans}
\calF_{ab}[\tilde\nabla] - \calF_{ab}[\nabla] = 2\nabla_{[a}\calA_{b]} - T_{ab}{}^c[\nabla] \calA_c + 2\calA_{[a} \calA_{b]}.
\end{equation}
This relation is a generalization of the transformation law~\eqref{FPreobr} for the case of non-symmetric connections.

Now we proceed to the derivation of Bianchi identities. For simplicity, we first assume that the connection $\nabla_a$ under consideration is symmetric. Consider the operator $2\nabla_{[a}\nabla_b\nabla_{c]}$, apply it to an arbitrary scalar $f$, and expand the resulting expression in two different ways. On one hand, using antisymmetry in indices $ab$ and the generalized Ricci identity \eqref{GenRicci}, and using antisymmetry in indices $bc$ and the definition of torsion \eqref{DefT} on the other hand:
\begin{align}
2\nabla_{[[a}\nabla_{b]}\nabla_{c]}f &= -R_{[abc]}{}^d\nabla_d f, \\
2\nabla_{[a}\nabla_{[b}\nabla_{c]]}f &= \nabla_{[a}T_{bc]}{}^d\nabla_d f.
\end{align}
Then, since the torsion is assumed to be zero $T_{ab}{}^c[\nabla] = 0$, we have $2\nabla_{[a}\nabla_b\nabla_{c]}f = 0$. Due to the fact that $\nabla_d f$ can take any values at an arbitrarily chosen point, we have the algebraic Bianchi identity \eqref{AlgBianSym}.

To obtain the differential Bianchi identity for the curvature in the bundle $\calF_{ab}$, we repeat the previous argument, but now we apply the operator $2\nabla_{[a}\nabla_b\nabla_{c]}$ not to the scalar, but to the quantity $\bm{\psi} \cong \psi^\alpha$:
\begin{align}
2\nabla_{[[a}\nabla_{b]}\nabla_{c]} \bm{\psi} &= \left(\calF_{[ab}\nabla_{c]} - R_{[abc]}{}^d\nabla_d\right) \bm{\psi} = \calF_{[ab}\nabla_{c]} \bm{\psi}, \\
2\nabla_{[a}\nabla_{[b}\nabla_{c]]} \bm{\psi} &= \nabla_{[a}\left(\calF_{bc]} \bm{\psi}\right) = \left((\nabla_{[a}\calF_{bc]}) + \calF_{[bc} \nabla_{a]}\right) \bm{\psi}.
\end{align}
In the first line, we took into account the algebraic identity \eqref{AlgBianSym}. Subtracting the first line from the second one we get $\left(\nabla_{[a}\calF_{bc]}\right)\bm{\psi} = 0$. Since $\bm{\psi}$ can take on any value at any point, we have $\nabla_{[a}\calF_{bc]} = 0$.

To obtain the transformation law and the differential Bianchi identity for the spacetime curvature $R_{abc}{}^d$, it suffices to simply repeat the corresponding arguments, replacing $\psi^\alpha$ with $v^c$, $\calA_a \cong \scrA_{a\alpha}{}^\beta$ with $\frakA_{a} \cong \scrA_{ab}{}^c$, and $\calF_{ab} \cong \scrF_{ab\alpha}{}^\beta$ with $\frakR_{ab} \cong R_{abc}{}^d$. Then we immediately obtain $\nabla_{[a}\frakR_{bc]} = 0$ and the following generalization of the relation \eqref{tildeFPreobr}:
\begin{equation} \label{genRtrans}
\frakR_{ab}[\tilde\nabla] - \frakR_{ab}[\nabla] = 2\nabla_{[a}\frakA_{b]} - T_{ab}{}^c[\nabla] \frakA_c + 2\frakA_{[a}\frakA_{b]}.
\end{equation}

We obtained the Bianchi identities \eqref{AlgBianSym}-\eqref{DifBian} for the symmetric connection. To obtain the corresponding generalizations to the case of connections with nonzero torsion $T_{ab}{}^c[\nabla] \ne0$, it suffices to simply write down the relations \eqref{AlgBianSym}-\eqref{DifBian} for the symmetric connection $\tilde\nabla_a$ and transform them according to \eqref{RaznDiff}, \eqref{ToSymmetric}, \eqref{genFtrans} and \eqref{genRtrans}. As a result, we get:
\begin{align}
&R_{[abc]}{}^d + \nabla_{[a} T_{bc]}{}^d + T_{[ab}{}^e T_{c]e}{}^d = 0, \label{genAlgBian} \\
&\nabla_{[a} \calF_{bc]} + T_{[ab}{}^d \calF_{c]d} = 0, \label{genDifBian1} \\
&\nabla_{[a} \frakR_{bc]} + T_{[ab}{}^d \frakR_{c]d} = 0. \label{genDifBian2}
\end{align}

%%%%%%%%%%%%%%%%%%%%%%%%%%%%%%%%%%%%%%%%%%%%%%%%%
\subsection{Matrix-valued differential forms} \label{FormApp}

Above we have already had enough reasons to see that it is extremely convenient to write potentials and curvatures as matrices $\calA_a \cong \scrA_{a\alpha}{}^\beta$, $\calF_{ab} \cong \scrF_{ab\alpha}{}^\beta$, $\frakA_a = \scrA_{ab}{}^c$, $\frakR_{ab} \cong R_{abc}{}^d$. On the other hand, both differential geometry and Yang-Mills theory have long and successfully used an alternative notation using differential forms. It turns out that these two approaches can be easily combined in the formalism of \emph{matrix-valued differential forms}. This notation allows us to write down the curvature transformation laws \eqref{genFtrans} and \eqref{genRtrans} and the differential Bianchi identities \eqref{genDifBian1}-\eqref{genDifBian2} in a very simple and elegant way. We have not used this formalism in the main text of the paper in order not to confuse the reader with possibly unusual notations. However, this approach is too beautiful to resist the temptation not to mention it at all and not to give the main definitions and results in this Appendix.

First of all, we recall that the usual (scalar-valued) differential $p$-forms are completely antisymmetric tensors with $p$ lower spacetime indices. We will replace these indices with the order $p$ in square brackets: $P_{[p]} \cong P_{i_1\ldots i_p} = P_{[i_1\ldots i_p]}$ etc. The \emph{exterior product} is the operation specified on such forms:
\begin{equation}
P_{[p]} \wedge Q_{[q]} \cong P_{[i_1\ldots i_p} Q_{i_{p+1}\ldots i_{p+q}]}.
\end{equation}
This operation is bilinear, associative and has the following graded commutation property
\begin{equation} \label{DifFormCommRel}
P_{[p]} \wedge Q_{[q]} = (-1)^{pq} Q_{[q]} \wedge P_{[p]}.
\end{equation}

By analogy, we define a matrix-valued differential $p$-form as a tensor that is simultaneously a matrix from the point of view of fibers and a $p$-form with respect to spacetime indices:
\begin{equation}
\bm{P}_{[p]} \cong P_{i_1\ldots i_p\alpha}{}^\beta = P_{[i_1\ldots i_p]\alpha}{}^\beta.
\end{equation}
Then their product is naturally defined as matrix multiplication with simultaneous antisymmetrization by indices of the form
\begin{equation}
\bm{P}_{[p]} \wedge \bm{Q}_{[q]} \cong P_{[i_1\ldots i_p|\gamma}{}^\beta Q_{|i_{p+1}\ldots i_{p+q }]\alpha}{}^\gamma.
\end{equation}
Obviously, the product of (matrix-valued) $p$- and $q$-forms is a (matrix-valued) $(p+q)$-form, and this operation is still bilinear and associative. However, due to the matrix structure, a commutation relation like \eqref{DifFormCommRel} no longer exists:
\begin{equation} \label{CommutatorDef}
[\bm{P}_{[p]}, \bm{Q}_{[q]}] = \bm{P}_{[p]} \wedge \bm{Q}_{[q]} - (-1)^{pq} \bm{Q}_{[q]} \wedge \bm{P}_{[p]} \ne 0.
\end{equation}

According to this definition, the potentials $\calA_{[1]} \cong \calA_a$ and $\frakA_{[1]} \cong \frakA_a$ are matrix-valued 1-forms, and the curvatures $\calF_{[2]} \cong \calF_{ab}$ and $\frakR_{[2]} \cong \frakR_{ab}$ are matrix-valued 2-forms.

Next, let $\nabla_a$ be a connection. We define the associated operation of \emph{exterior derivative} $d[\nabla]$ using the relation
\begin{equation} \label{ExtDifDef}
d[\nabla]\bm{P}_{[p]} \cong \nabla_{[i_1} \bm{P}_{i_2\ldots i_{p+1}]} - \frac{p}{2}T_{[i_1i_2}{}^a \bm{P}_{a|i_3\ldots i_{p+1}]}.
\end{equation}
Obviously, this operation transforms the (matrix-valued) $p$-form into a (matrix-valued) $(p+1)$-form, it is linear and, moreover, is an odd quantity in the sense that it satisfies the following Leibniz rule:
\begin{equation}
d \left(\bm{P}_{[p]}\wedge \bm{Q}_{[q]}\right) = \left(d\bm{P}_{[p]}\right) \wedge \bm{Q}_{[q]} + (-1)^p \bm{P}_{[p]} \wedge \left(d \bm{Q}_{[q]}\right).
\end{equation}

The exterior derivative defined in this way, in general, depends on the connection $\nabla_a$, but it has the following remarkable property: the connection transformations associated with form indices exactly cancel with the torsion transformation, leaving only the part associated with the matrix indices, elegantly expressed in terms of the commutator with potential (here $\bm{P}_{[p]}\cong P_{[p]\alpha}{}^\beta$ and $\frakP_{[p]} \cong P_{[p]a}{}^b$):
\begin{align} 
&\Big(d[\tilde\nabla] - d[\nabla]\Big) \bm{P}_{[p]} = [\calA_{[1]}, \bm{P}_{[p]}], \label{ExtDerTrans} \\
&\Big(d[\tilde\nabla] - d[\nabla]\Big) \frakP_{[p]} = [\frakA_{[1]}, \frakP_{[p]}].
\end{align}
In the special case of scalar-valued $p$-forms, the commutator on the right side of the transformation \eqref{ExtDerTrans} disappears, and we arrive at the famous result that the exterior derivative $d$ does not depend at all on the connection $\nabla_a$ and, therefore, it is an operation determined exclusively by the differential structure of the manifold.

Another remarkable property of the exterior derivative is that the action of its square on an arbitrary matrix-valued $p$-form is given by its commutator with curvature 2-form:
\begin{equation} \label{ExtDerSquareEq}
d^2 \bm{P}_{[p]} = \frac{1}{2}[\calF_{[2]}, \bm{P}_{[p]}],  \quad d^2 \frakP_{[p]} = \frac{1}{2}[\frakR_{[2]}, \frakP_{[p]}].
\end{equation}
Again, in the case of scalar-valued $p$-forms, the commutator on the right disappears, and we obtain the well-known relation
\begin{equation}
d^2 = 0.
\end{equation}

Finally, we use this formalism to write down the curvature transformation laws \eqref{genFtrans} and \eqref{genRtrans}
\begin{align}
&\calF_{[2]}[\tilde\nabla] - \calF_{[2]}[\nabla] = 2d[\nabla] \calA_{[1]} + [\calA_{[1]}, \calA_{[1]}], \label{genFtrans2} \\
&\frakR_{[2]}[\tilde\nabla] - \frakR_{[2]}[\nabla] = 2d[\nabla] \frakA_{[1]} + [\frakA_{[1]} , \frakA_{[1]}], \label{genRtrans2}
\end{align}
and the differential Bianchi identities \eqref{genDifBian1}-\eqref{genDifBian2}
\begin{equation}
d\calF_{[2]} = 0, \qquad d\frakR_{[2]} = 0.
\end{equation}

%%%%%%%%%%%%%%%%%%%%%%%%%%%%%%%%%%%%%%%%%%%%%%%%%
\section{An arbitrary given background} \label{VertApp}
%%%%%%%%%%%%%%%%%%%%%%%%%%%%%%%%%%%%%%%%%%%%%%%%%

In this Appendix, for reference, we provide expressions for an arbitrary nontrivial background field configuration given by $\bm{F}_{ab}$, $\bm{N}_a$.

\subsection{Interaction vertices}

To obtain all interaction vertices, we need to take the expressions \eqref{CalL00h}-\eqref{CalL22h}, replace $\partial_a$ with covariant derivatives $\nabla_a$, which no longer commute with each other or with the operation of Hermitian conjugation, and add the corrections given below. As previously in Subsection~\ref{VerticesWithH}, we do not expand the matrices $\bm{\omega}$ and $\bm{\Omega}$ \eqref{OomegaNewDef} into a power series in $\bm{h}$. For brevity, we also use the notation $e_\pm^2$ \eqref{Epm}.

Fourth-order vertices in the fields $\bm{A}_a$ and $\bm{B}_a$ do not receive any corrections from the background fields. Corrections for the third-order vertices will have the form:
\begin{align}
\Delta\calL_{30} &= -\frac{ie}{4\tilde e} \tr\big(e_+^2 \hat{\bm C}_{ab}\bm{\Omega}\hat{\bm K}^{ab}\bm{\omega} + e_-^2 \hat{\bm C}_{ab}\hat{\bm K}^{ab}\big), \\
\Delta\calL_{03} &= \frac{\tilde e^2}{4e^2} \tr\big(e_+^2 \check{\bm C}_{ab}\bm{\Omega}\check{\bm K}^{ab}\bm{\omega} + e_-^2 \check{\bm C}_{ab}\check{\bm K}^{ab}\big), \\
\Delta\calL_{21} &= \frac{1}{4} \tr\Big(e_+^2 \big(\bm{C}_{ab}\bm{\Omega}\hat{\bm K}^{ab}\bm{\omega} - \hat{\bm C}_{ab}\bm{\Omega}\check{\bm K}^{ab}\bm{\omega}\big) \nonumber \\
&- e_-^2 \big(\bm{C}_{ab}\hat{\bm K}^{ab} + \hat{\bm C}_{ab}\check{\bm K}^{ab}\big)\Big), \\
\Delta\calL_{12} &= \frac{i\tilde e}{4e} \tr\Big(e_+^2 \big(\check{\bm C}_{ab}\bm{\Omega}\hat{\bm K}^{ab}\bm{\omega} - \bm{C}_{ab}\bm{\Omega}\check{\bm K}^{ab}\bm{\omega}\big) \nonumber \\
&+ e_-^2\big(\check{\bm C}_{ab}\hat{\bm K}^{ab} + \bm{C}_{ab}\check{\bm K}^{ab}\big) \Big).
\end{align}

For the second-order corrections in the fields $\bm{A}_a$ and $\bm{B}_a$, we have:
\begin{align}
\Delta\calL_{20} &= \frac{i}{4\tilde e} \tr\big(e_+^2 \hat{\bm D}_{ab}\bm{\Omega}\hat{\bm K}^{ab}\bm{\omega} + e_-^2 \hat{\bm D}_{ab}\hat{\bm K}^{ab}\big) \nonumber \\
&+ \frac{ie_+^2}{8\tilde e}\tr\big(\hat{\bm C}_{ab}\bm{\Omega}\bm{G}^{ab}\bm{\omega} - \bm{G}_{ab}\bm{\Omega}\hat{\bm C}^{ab}\bm{\omega}\big) \nonumber \\
&- \frac{e}{8\tilde e^2}\tr\Big(e_+^2 \big(\hat{\bm C}_{ab}\bm{\Omega}\bm{F}^{ab}\bm{\omega} + \bm{F}_{ab}\bm{\Omega}\hat{\bm C}^{ab}\bm{\omega}\big) \nonumber \\
&+ 2e_-^2 \big(\hat{\bm C}_{ab}\bm{F}^{ab}\big)\Big) - \frac{e_-^2}{4}\tr\big(\hat{\bm K}_{ab}\hat{\bm K}^{ab}\big), \\
\Delta\calL_{02} &= \frac{i\tilde e}{4e^2} \tr\big(e_+^2 \check{\bm D}_{ab}\bm{\Omega}\check{\bm K}^{ab}\bm{\omega} + e_-^2 \check{\bm D}_{ab}\check{\bm K}^{ab}\big) \nonumber \\
&- \frac{i\tilde ee_+^2}{8e^2}\tr\big(\check{\bm C}_{ab}\bm{\Omega}\bm{G}^{ab}\bm{\omega} - \bm{G}_{ab}\bm{\Omega}\check{\bm C}^{ab}\bm{\omega}\big) \nonumber \\
&+ \frac{1}{8e}\tr\Big(e_+^2 \big(\check{\bm C}_{ab}\bm{\Omega}\bm{F}^{ab}\bm{\omega} + \bm{F}_{ab}\bm{\Omega}\check{\bm C}^{ab}\bm{\omega}\big) \nonumber \\
&+ 2e_-^2 \big(\check{\bm C}_{ab}\bm{F}^{ab}\big)\Big) + \frac{\tilde e^2e_-^2}{4e^2}\tr\big(\check{\bm K}_{ab}\check{\bm K}^{ab}\big), \\
\Delta\calL_{11} &= \frac{1}{4e} \tr\Big(e_+^2 \big(\hat{\bm D}_{ab}\bm{\Omega}\check{\bm K}^{ab}\bm{\omega} - \check{\bm D}_{ab}\bm{\Omega}\hat{\bm K}^{ab}\bm{\omega}\big) \nonumber \\
&+ e_-^2 \big(\hat{\bm D}_{ab}\check{\bm K}^{ab} + \check{\bm D}_{ab}\hat{\bm K}^{ab}\big)\Big) + \frac{1}{8e}\tr\Big(e_-^2 \bm{C}_{ab}\bm{G}^{ab} \nonumber \\
&- e_+^2\big(\bm{C}_{ab}\bm{\Omega}\bm{G}^{ab}\bm{\omega} + \bm{G}_{ab}\bm{\Omega}\bm{C}^{ab}\bm{\omega}\big)\Big)\nonumber \\
&- \frac{ie_+^2}{8\tilde e}\tr\big(\bm{C}_{ab}\bm{\Omega}\bm{F}^{ab}\bm{\omega} - \bm{F}_{ab}\bm{\Omega}\bm{C}^{ab}\bm{\omega}\big) \nonumber \\
&+ \frac{i\tilde ee_-^2}{2e}\tr\big(\check{\bm K}_{ab}\hat{\bm K}^{ab}\big).
\end{align}

For terms linear in $\bm{A}_a$ and $\bm{B}_a$, we obtain the following expressions:
\begin{align}
\Delta\calL_{10} &= \frac{ieM^2}{2\tilde e}\tr\big(\bm{A}_a\bm{N}^a - \bm{A}_a\bm{\Omega}\bm{N}^a\bm{\omega}\big) \nonumber \\
&+ \frac{ie_+^2}{8e\tilde e} \tr\big(\bm{G}_{ab}\bm{\Omega}\hat{\bm D}^{ab}\bm{\omega} - \hat{\bm D}_{ab}\bm{\Omega}\bm{G}^{ab}\bm{\omega}\big) \nonumber \\
&+ \frac{1}{4e}\tr\big(e_-^2\bm{G}_{ab}\hat{\bm K}^{ab} - e_+^2\bm{G}_{ab}\bm{\Omega}\hat{\bm K}^{ab}\bm{\omega}\big) \nonumber \\
&+ \frac{i}{4\tilde e}\tr\big(e_+^2 \bm{F}_{ab}\bm{\Omega}\hat{\bm K}^{ab}\bm{\omega} + e_-^2 \bm{F}_{ab}\hat{\bm K}^{ab}\big) \nonumber \\
&+ \frac{1}{8\tilde e^2}\tr\Big(e_+^2 \big(\hat{\bm D}_{ab}\bm{\Omega}\bm{F}^{ab}\bm{\omega} + \bm{F}_{ab}\bm{\Omega}\hat{\bm D}^{ab}\bm{\omega}\big) \nonumber \\
&+ 2e_-^2\hat{\bm D}_{ab}\bm{F}^{ab}\Big) \\
\Delta\calL_{01} &= \frac{M^2}{2}\tr\big(\bm{B}_a\bm{N}^a + \bm{B}_a\bm{\Omega}\bm{N}^a\bm{\omega}\big) \nonumber \\
&+ \frac{ie_+^2}{8e\tilde e} \tr\big(\check{\bm D}_{ab}\bm{\Omega}\bm{F}^{ab}\bm{\omega} - \bm{F}_{ab}\bm{\Omega}\check{\bm D}^{ab}\bm{\omega}\big) \nonumber \\
&+ \frac{1}{4e}\tr\big(e_-^2\bm{F}_{ab}\check{\bm K}^{ab} + e_+^2\bm{F}_{ab}\bm{\Omega}\check{\bm K}^{ab}\bm{\omega}\big) \nonumber \\
&+ \frac{i\tilde e}{4e^2}\tr\big(e_+^2 \bm{G}_{ab}\bm{\Omega}\check{\bm K}^{ab}\bm{\omega} - e_-^2 \bm{G}_{ab}\check{\bm K}^{ab}\big) \nonumber \\
&+ \frac{1}{8e^2}\tr\Big(e_+^2 \big(\check{\bm D}_{ab}\bm{\Omega}\bm{G}^{ab}\bm{\omega} + \bm{G}_{ab}\bm{\Omega}\check{\bm D}^{ab}\bm{\omega}\big) \nonumber \\
&- 2e_-^2\check{\bm D}_{ab}\bm{G}^{ab}\Big).
\end{align}

Finally, the corrections that do not contain $\bm{A}_a$ and $\bm{B}_a$ will have the form:
\begin{align}
\calL_{00} &= \frac{e_+^2}{8}\tr\Big(\frac{1}{\tilde e^2} \big(\bm{F}_{ab}\bm{\Omega}\bm{F}^{ab}\bm{\omega}) + \frac{1}{e^2} \big(\bm{G}_{ab}\bm{\Omega}\bm{G}^{ab}\bm{\omega} \big) \nonumber \\
&+ \frac{i}{e\tilde e} \big(\bm{G}_{ab}\bm{\Omega}\bm{F}^{ab}\bm{\omega} - \bm{F}_{ab}\bm{\Omega}\bm{G}^{ab}\bm{\omega}\big)\Big) \nonumber \\
&- \frac{M^2}{2\tilde e} \tr\big(\bm{\Omega}\bm{N}^a\nabla_a\bm{\omega}\big).
\end{align}

%%%%%%%%%%%%%%%%%%%%%%%%%%%%%%%%%%%%%%%%%%%%%%%%%
\subsection{Linearized EoMs} \label{PropApp}

In particular, for the part of the Lagrangian that is quadratic in small perturbations of $\bm{A}_a$, $\bm{B}_a$, and $\bm{h}$, we obtain the following corrections to expressions \eqref{TrivLinAction}-\eqref{TrivLinAction2}:
\begin{align}
\Delta\calL_{AA}^{(2)} &= \tr\Big(\frac{i\tilde e}{2}\hat{\bm D}_{ab}\hat{\bm K}^{ab} -\frac{e}{2}\bm{F}_{ab}\hat{\bm C}^{ab} - \frac{e_-^2}{4}\hat{\bm K}_{ab}\hat{\bm K}^{ab}\Big), \\
\Delta\calL_{BB}^{(2)} &= \tr\Big(\frac{i\tilde e}{2}\check{\bm D}_{ab}\check{\bm K}^{ab} + \frac{\tilde e^2}{2e}\bm{F}_{ab}\check{\bm C}^{ab} + \frac{\tilde e^2 e_-^2}{4e^2}\check{\bm K}_{ab}\check{\bm K}^{ab}\Big), \\
\Delta\calL_{AB}^{(2)} &= \tr\Big(\frac{\tilde e^2}{2e}\check{\bm K}_{ab}\hat{\bm D}^{ab} - \frac{e}{2}\hat{\bm K}_{ab}\check{\bm D}^{ab} -\frac{e}{2}\bm{G}_{ab}\bm{C}^{ab} \nonumber \\
&+ \frac{i\tilde e e_-^2}{2e}\check{\bm K}_{ab}\hat{\bm K}^{ab}\Big),
\end{align}
\begin{align}
\Delta\calL_{hh}^{(2)} &= \tilde e \tr\big([\bm{h}, \bm{N}^a]\nabla_a\bm{h}\big) + \frac{e_+^2}{2M^2} \tr\Big( [\bm{h}, \bm{F}_{ab}]\bm{F}^{ab}\bm{h} \nonumber \\
&+ \frac{\tilde e^2}{e^2}[\bm{h}, \bm{G}_{ab}]\bm{G}^{ab}\bm{h} \Big), \\
\Delta\calL_{Ah}^{(2)} &= ieM\tr\big([\bm{N}_a, \bm{A}^a]\bm{h}\big) + \frac{e_+^2}{2M} \tr\Big(\frac{i}{e} [\bm{G}_{ab}, \hat{\bm D}^{ab}]\bm{h} \nonumber \\
&+ i [\bm{F}_{ab}, \hat{\bm K}^{ab}]\bm{h} + \frac{\tilde e}{e}[\hat{\bm K}_{ab}, \bm{G}^{ab}]\bm{h} \Big), \\
\Delta\calL_{Bh}^{(2)} &= \tilde eM \tr\big([\bm{B}_a, \bm{N}^a]\bm{h}\big) + \frac{e_+^2}{2M} \tr\Big(\frac{i}{e} [\check{\bm D}_{ab}, \bm{F}^{ab}]\bm{h} \nonumber \\
&+ \frac{\tilde e}{e} [\bm{F}_{ab}, \check{\bm K}^{ab}]\bm{h} + \frac{i\tilde e^2}{e^2}[\bm{G}_{ab}, \check{\bm K}_{ab}]\bm{h} \Big).
\end{align}

Using the notation
\begin{equation}
H_{IJ} \phi_J = \frac{\delta\calL_{IJ}^{(2)}}{\delta\phi_I},
\end{equation}
the system of linearized EoMs can be rewritten as
\begin{equation}
\begin{pmatrix}
H_{AA} & H_{AB} & H_{Ah} \\
H_{BA} & H_{BB} & H_{Bh} \\
H_{hA} & H_{hB} & H_{hh}
\end{pmatrix}
\begin{pmatrix} \bm{A}_a \\ \bm{B}_a \\ \bm{h} \end{pmatrix}
=0.
\end{equation}

Further, all nine variations are found without difficulty, although they yield rather cumbersome expressions. Varying over $\bm{A}_a$ we get:
\begin{align}
\frac{\calL_{AA}^{(2)}}{\delta\bm{A}_a} &= \nabla_b\hat{\bm D}^{ab} + \tilde e[\hat{\bm D}^{ab}, \bm{N}_b] + i\tilde e\nabla_b\hat{\bm K}^{ab} \nonumber \\
&+ ie[\bm{F}^{ab}, \bm{A}_b] + ie_-^2[\hat{\bm K}^{ab}, \bm{N}_b], \\
\frac{\calL_{AB}^{(2)}}{\delta\bm{A}_a} &= ie[\check{\bm D}^{ab}, \bm{N}_b] + \frac{\tilde e^2}{e}\nabla_b\check{\bm K}^{ab} \nonumber \\
&+ ie[\bm{G}^{ab}, \bm{B}_b] + \frac{\tilde ee_-^2}{e}[\check{\bm K}^{ab}, \bm{N}_b], \\
\frac{\calL_{Ah}^{(2)}}{\delta\bm{A}_a} &= ieM[\bm{h}, \bm{N}^a] + \frac{e_+^2}{M} \Big(\frac{i}{e} \nabla_b[\bm{h}, \bm{G}^{ab}] \nonumber \\
&+ \big[[\bm{h}, \bm{F}^{ab}], \bm{N}_b\big] + \frac{i\tilde e}{e} \big[[\bm{h}, \bm{G}^{ab}], \bm{N}_b\big] \Big).
\end{align}
Varying over $\bm{B}_a$ yields:
\begin{align}
\frac{\calL_{BB}^{(2)}}{\delta\bm{B}_a} &= \nabla_b\hat{\bm D}^{ab} + M^2\bm{B}^a + \tilde e[\check{\bm D}^{ab}, \bm{N}_b] + i\tilde e\nabla_b\check{\bm K}^{ab} \nonumber \\
&- \frac{i\tilde e^2}{e} [\bm{F}^{ab}, \bm{B}_b] - \frac{i\tilde e^2e_-^2}{e^2} [\check{\bm K}^{ab}, \bm{N}_b], \\
\frac{\calL_{AB}^{(2)}}{\delta\bm{B}_a} &= -\frac{i\tilde e^2}{e} [\hat{\bm D}^{ab}, \bm{N}_b] + e\nabla_b\hat{\bm K}^{ab} \nonumber \\
&+ ie[\bm{G}^{ab}, \bm{A}_b] + \frac{\tilde ee_-^2}{e}[\hat{\bm K}^{ab}, \bm{N}_b], \\
\frac{\calL_{Bh}^{(2)}}{\delta\bm{B}_a} &= -M\nabla^a\bm{h} - \tilde eM[\bm{h}, \bm{N}^a] + \frac{e_+^2}{M} \Big(-\frac{i}{e} \nabla_b[\bm{h}, \bm{F}^{ab}] \nonumber \\
&- \frac{i\tilde e}{e} \big[[\bm{h}, \bm{F}^{ab}], \bm{N}_b\big] + \frac{\tilde e^2}{e^2} \big[[\bm{h}, \bm{G}^{ab}], \bm{N}_b\big] \Big).
\end{align}
Finally, varying with respect to $\bm{h}$ gives us the last three terms:
\begin{align}
\frac{\calL_{hh}^{(2)}}{\delta\bm{h}} &= \Box\bm{h} + \tilde e \big(2[\bm{N}^a, \nabla_a\bm{h}] + [\nabla_a\bm{N}^a, \bm{h}]\big) \nonumber \\
&+ \frac{e_+^2}{2M^2}\Big(\big[[\bm{h}, \bm{F}_{ab}], \bm{F}^{ab}\big] + \frac{\tilde e^2}{e^2} \big[[\bm{h}, \bm{G}_{ab}], \bm{G}^{ab}\big] \Big), \\
\frac{\calL_{Ah}^{(2)}}{\delta\bm{h}} &= ieM [\bm{N}_a, \bm{A}^a] + \frac{e_+^2}{2M} \Big(-\frac{i}{e}[\hat{\bm D}_{ab}, \bm{G}^{ab}] \nonumber \\
&+i[\bm{F}_{ab}, \hat{\bm K}^{ab}] - \frac{\tilde e}{e}[\bm{G}_{ab}, \hat{\bm K}^{ab}]\Big), \\
\frac{\calL_{Bh}^{(2)}}{\delta\bm{h}} &= -\tilde eM [\bm{N}_a, \bm{B}^a] + \frac{e_+^2}{2M} \Big(\frac{i}{e}[\hat{\bm D}_{ab}, \bm{F}^{ab}] \nonumber \\
&+ \frac{\tilde e}{e} [\bm{F}_{ab}, \check{\bm K}^{ab}] + \frac{i\tilde e^2}{e^2} [\bm{G}_{ab}, \check{\bm K}^{ab}]\Big).
\end{align}

\bibliographystyle{apsrev4-2}
\bibliography{Wachowski2604}

\end{document}